\documentclass[aps,prl,twocolumn,superscriptaddress,preprintnumbers,floatfix]{revtex4-1}

\usepackage{longtable} 
\usepackage{graphicx}
\usepackage{amsmath}
\usepackage{amssymb}
\usepackage{mdwlist}
\usepackage[caption=false]{subfig}
\usepackage{siunitx}
\usepackage{placeins}
\usepackage{color}
\usepackage[colorlinks=True]{hyperref}
\usepackage{standalone}
\usepackage{dcolumn}
\usepackage{microtype}
\usepackage{gensymb}
\usepackage{etoolbox}

\AtBeginDocument{\usepackage{booktabs}}

\newcommand{\ts}{\textsuperscript}

\newcommand{\beq}{\begin{equation}}
\newcommand{\eeq}{\end{equation}}

\newcommand*{\eq}[1]{Eq.\ (\ref{eq:#1})}
\newcommand*{\fig}[1]{Fig.\ \ref{fig:#1}}

\newcommand{\frot}{f_{\rm rot}}
\newcommand{\fgw}{f_{\rm GW}}

\newcommand{\hul}[1]{h_{\rm #1}^{95\%}}
\newcommand{\hT}{{h_{\rm t}}}
\newcommand{\hV}{{h_{\rm v}}}
\newcommand{\hS}{{h_{\rm s}}}

\newcommand{\data}{{\bf B}}

\newcommand*{\hyp}[1]{{\cal H}_{\rm #1}}
\newcommand{\hypn}{\hyp{N}}
\newcommand{\hyps}{\hyp{S}}

\newcommand*{\bayes}[2]{{\cal B}^{\rm #1}_{\rm #2}}
\newcommand{\bayessn}{\bayes{S}{N}}

\newcommand*{\odds}[2]{{\cal O}^{\rm #1}_{\rm #2}}
\newcommand{\oddssn}{\odds{S}{N}}
\newcommand{\oddsci}{\odds{S}{I}}

\newcommand{\ndet}{D}

\newcommand{\p}{P}

\newcommand*{\red}[1]{#1} 
\newcommand*{\blue}[1]{#1} 

\newcommand{\npsrs}{200}

\newcommand{\header}[1]{{\bf \em #1}}

\newtoggle{fullauthorlist}
\toggletrue{fullauthorlist}

\newtoggle{endauthorlist}
\toggletrue{endauthorlist}

\graphicspath{{./fig/}}

\begin{document}

\preprint{LIGO-P1700009}

\title{First search for nontensorial gravitational waves from known pulsars}



\iftoggle{endauthorlist}{
 %
 %
 \let\mymaketitle\maketitle
 \let\myauthor\author
 \let\myaffiliation\affiliation
 \author{The LIGO Scientific Collaboration and the Virgo Collaboration, \\
S.~Buchner,
I.~Cognard,
A.~Corongiu,
P.~C.~C.~Freire,
L.~Guillemot,
G.~B.~Hobbs,
M.~Kerr,
A.~G.~Lyne,
A.~Possenti,
A.~Ridolfi,
R.~M.~Shannon,
B.~W.~Stappers,
and P.~Weltevrede}
}{
 %
 %
 \iftoggle{fullauthorlist}{

\author{%
B.~P.~Abbott,$^{1}$  
R.~Abbott,$^{1}$  
T.~D.~Abbott,$^{2}$  
F.~Acernese,$^{3,4}$ 
K.~Ackley,$^{5}$  
C.~Adams,$^{6}$  
T.~Adams,$^{7}$ 
P.~Addesso,$^{8}$  
R.~X.~Adhikari,$^{1}$  
V.~B.~Adya,$^{9}$  
C.~Affeldt,$^{9}$  
M.~Afrough,$^{10}$  
B.~Agarwal,$^{11}$  
M.~Agathos,$^{12}$	
K.~Agatsuma,$^{13}$ 
N.~Aggarwal,$^{14}$  
O.~D.~Aguiar,$^{15}$  
L.~Aiello,$^{16,17}$ 
A.~Ain,$^{18}$  
P.~Ajith,$^{19}$  
G.~Allen,$^{11}$  
A.~Allocca,$^{20,21}$ 
P.~A.~Altin,$^{22}$  
A.~Amato,$^{23}$ %
A.~Ananyeva,$^{1}$  
S.~B.~Anderson,$^{1}$  
W.~G.~Anderson,$^{24}$  
S.~Antier,$^{25}$ 
S.~Appert,$^{1}$  
K.~Arai,$^{1}$	
M.~C.~Araya,$^{1}$  
J.~S.~Areeda,$^{26}$  
N.~Arnaud,$^{25,27}$ 
K.~G.~Arun,$^{28}$  
S.~Ascenzi,$^{29,17}$ 
G.~Ashton,$^{9}$  
M.~Ast,$^{30}$  
S.~M.~Aston,$^{6}$  
P.~Astone,$^{31}$ 
P.~Aufmuth,$^{32}$  
C.~Aulbert,$^{9}$  
K.~AultONeal,$^{33}$  
A.~Avila-Alvarez,$^{26}$  
S.~Babak,$^{34}$  
P.~Bacon,$^{35}$ 
M.~K.~M.~Bader,$^{13}$ 
S.~Bae,$^{36}$  
P.~T.~Baker,$^{37,38}$  
F.~Baldaccini,$^{39,40}$ 
G.~Ballardin,$^{27}$ 
S.~W.~Ballmer,$^{41}$  
S.~Banagiri,$^{42}$  
J.~C.~Barayoga,$^{1}$  
S.~E.~Barclay,$^{43}$  
B.~C.~Barish,$^{1}$  
D.~Barker,$^{44}$  
F.~Barone,$^{3,4}$ 
B.~Barr,$^{43}$  
L.~Barsotti,$^{14}$  
M.~Barsuglia,$^{35}$ 
D.~Barta,$^{45}$ 
J.~Bartlett,$^{44}$  
I.~Bartos,$^{46}$  
R.~Bassiri,$^{47}$  
A.~Basti,$^{20,21}$ 
J.~C.~Batch,$^{44}$  
C.~Baune,$^{9}$  
M.~Bawaj,$^{48,40}$ %
M.~Bazzan,$^{49,50}$ 
B.~B\'ecsy,$^{51}$  
C.~Beer,$^{9}$  
M.~Bejger,$^{52}$ 
I.~Belahcene,$^{25}$ 
A.~S.~Bell,$^{43}$  
B.~K.~Berger,$^{1}$  
G.~Bergmann,$^{9}$  
C.~P.~L.~Berry,$^{53}$  
D.~Bersanetti,$^{54,55}$ 
A.~Bertolini,$^{13}$ 
J.~Betzwieser,$^{6}$  
S.~Bhagwat,$^{41}$  
R.~Bhandare,$^{56}$  
I.~A.~Bilenko,$^{57}$  
G.~Billingsley,$^{1}$  
C.~R.~Billman,$^{5}$  
J.~Birch,$^{6}$  
R.~Birney,$^{58}$  
O.~Birnholtz,$^{9}$  
S.~Biscans,$^{14}$  
A.~Bisht,$^{32}$  
M.~Bitossi,$^{27,21}$ 
C.~Biwer,$^{41}$  
M.~A.~Bizouard,$^{25}$ 
J.~K.~Blackburn,$^{1}$  
J.~Blackman,$^{59}$  
C.~D.~Blair,$^{60}$  
D.~G.~Blair,$^{60}$  
R.~M.~Blair,$^{44}$  
S.~Bloemen,$^{61}$ 
O.~Bock,$^{9}$  
N.~Bode,$^{9}$  
M.~Boer,$^{62}$ 
G.~Bogaert,$^{62}$ 
A.~Bohe,$^{34}$  
F.~Bondu,$^{63}$ 
R.~Bonnand,$^{7}$ 
B.~A.~Boom,$^{13}$ 
R.~Bork,$^{1}$  
V.~Boschi,$^{20,21}$ 
S.~Bose,$^{64,18}$  
Y.~Bouffanais,$^{35}$ 
A.~Bozzi,$^{27}$ 
C.~Bradaschia,$^{21}$ 
P.~R.~Brady,$^{24}$  
V.~B.~Braginsky$^*$,$^{57}$  
M.~Branchesi,$^{65,66}$ 
J.~E.~Brau,$^{67}$   
T.~Briant,$^{68}$ 
A.~Brillet,$^{62}$ 
M.~Brinkmann,$^{9}$  
V.~Brisson,$^{25}$ 
P.~Brockill,$^{24}$  
J.~E.~Broida,$^{69}$  
A.~F.~Brooks,$^{1}$  
D.~A.~Brown,$^{41}$  
D.~D.~Brown,$^{53}$  
N.~M.~Brown,$^{14}$  
S.~Brunett,$^{1}$  
C.~C.~Buchanan,$^{2}$  
A.~Buikema,$^{14}$  
T.~Bulik,$^{70}$ 
H.~J.~Bulten,$^{71,13}$ 
A.~Buonanno,$^{34,72}$  
D.~Buskulic,$^{7}$ 
C.~Buy,$^{35}$ 
R.~L.~Byer,$^{47}$ 
M.~Cabero,$^{9}$  
L.~Cadonati,$^{73}$  
G.~Cagnoli,$^{23,74}$ 
C.~Cahillane,$^{1}$  
J.~Calder\'on~Bustillo,$^{73}$  
T.~A.~Callister,$^{1}$  
E.~Calloni,$^{75,4}$ 
J.~B.~Camp,$^{76}$  
M.~Canepa,$^{54,55}$ 
P.~Canizares,$^{61}$ 
K.~C.~Cannon,$^{77}$  
H.~Cao,$^{78}$  
J.~Cao,$^{79}$  
C.~D.~Capano,$^{9}$  
E.~Capocasa,$^{35}$ 
F.~Carbognani,$^{27}$ 
S.~Caride,$^{80}$  
M.~F.~Carney,$^{81}$  
J.~Casanueva~Diaz,$^{25}$ 
C.~Casentini,$^{29,17}$ 
S.~Caudill,$^{24}$  
M.~Cavagli\`a,$^{10}$  
F.~Cavalier,$^{25}$ 
R.~Cavalieri,$^{27}$ 
G.~Cella,$^{21}$ 
C.~B.~Cepeda,$^{1}$  
L.~Cerboni~Baiardi,$^{65,66}$ 
G.~Cerretani,$^{20,21}$ 
E.~Cesarini,$^{29,17}$ 
S.~J.~Chamberlin,$^{82}$  
M.~Chan,$^{43}$  
S.~Chao,$^{83}$  
P.~Charlton,$^{84}$  
E.~Chassande-Mottin,$^{35}$ 
D.~Chatterjee,$^{24}$  
B.~D.~Cheeseboro,$^{37,38}$  
H.~Y.~Chen,$^{85}$  
Y.~Chen,$^{59}$  
H.-P.~Cheng,$^{5}$  
A.~Chincarini,$^{55}$ 
A.~Chiummo,$^{27}$ 
T.~Chmiel,$^{81}$  
H.~S.~Cho,$^{86}$  
M.~Cho,$^{72}$  
J.~H.~Chow,$^{22}$  
N.~Christensen,$^{69,62}$  
Q.~Chu,$^{60}$  
A.~J.~K.~Chua,$^{12}$  
S.~Chua,$^{68}$ 
A.~K.~W.~Chung,$^{87}$  
S.~Chung,$^{60}$  
G.~Ciani,$^{5}$  
R.~Ciolfi,$^{88,89}$ 
C.~E.~Cirelli,$^{47}$  
A.~Cirone,$^{54,55}$ 
F.~Clara,$^{44}$  
J.~A.~Clark,$^{73}$  
F.~Cleva,$^{62}$ 
C.~Cocchieri,$^{10}$  
E.~Coccia,$^{16,17}$ 
P.-F.~Cohadon,$^{68}$ 
A.~Colla,$^{90,31}$ 
C.~G.~Collette,$^{91}$  
L.~R.~Cominsky,$^{92}$  
M.~Constancio~Jr.,$^{15}$  
L.~Conti,$^{50}$ 
S.~J.~Cooper,$^{53}$  
P.~Corban,$^{6}$  
T.~R.~Corbitt,$^{2}$  
K.~R.~Corley,$^{46}$  
N.~Cornish,$^{93}$  
A.~Corsi,$^{80}$  
S.~Cortese,$^{27}$ 
C.~A.~Costa,$^{15}$  
M.~W.~Coughlin,$^{69}$  
S.~B.~Coughlin,$^{94,95}$  
J.-P.~Coulon,$^{62}$ 
S.~T.~Countryman,$^{46}$  
P.~Couvares,$^{1}$  
P.~B.~Covas,$^{96}$  
E.~E.~Cowan,$^{73}$  
D.~M.~Coward,$^{60}$  
M.~J.~Cowart,$^{6}$  
D.~C.~Coyne,$^{1}$  
R.~Coyne,$^{80}$  
J.~D.~E.~Creighton,$^{24}$  
T.~D.~Creighton,$^{97}$  
J.~Cripe,$^{2}$  
S.~G.~Crowder,$^{98}$  
T.~J.~Cullen,$^{26}$  
A.~Cumming,$^{43}$  
L.~Cunningham,$^{43}$  
E.~Cuoco,$^{27}$ 
T.~Dal~Canton,$^{76}$  
S.~L.~Danilishin,$^{32,9}$  
S.~D'Antonio,$^{17}$ 
K.~Danzmann,$^{32,9}$  
A.~Dasgupta,$^{99}$  
C.~F.~Da~Silva~Costa,$^{5}$  
V.~Dattilo,$^{27}$ 
I.~Dave,$^{56}$  
M.~Davier,$^{25}$ 
D.~Davis,$^{41}$  
E.~J.~Daw,$^{100}$  
B.~Day,$^{73}$  
S.~De,$^{41}$  
D.~DeBra,$^{47}$  
J.~Degallaix,$^{23}$ 
M.~De~Laurentis,$^{75,4}$ 
S.~Del\'eglise,$^{68}$ 
W.~Del~Pozzo,$^{53,20,21}$ 
T.~Denker,$^{9}$  
T.~Dent,$^{9}$  
V.~Dergachev,$^{34}$  
R.~De~Rosa,$^{75,4}$ 
R.~T.~DeRosa,$^{6}$  
R.~DeSalvo,$^{101}$  
J.~Devenson,$^{58}$  
R.~C.~Devine,$^{37,38}$  
S.~Dhurandhar,$^{18}$  
M.~C.~D\'{\i}az,$^{97}$  
L.~Di~Fiore,$^{4}$ 
M.~Di~Giovanni,$^{102,89}$ 
T.~Di~Girolamo,$^{75,4,46}$ 
A.~Di~Lieto,$^{20,21}$ 
S.~Di~Pace,$^{90,31}$ 
I.~Di~Palma,$^{90,31}$ 
F.~Di~Renzo,$^{20,21}$ %
Z.~Doctor,$^{85}$  
V.~Dolique,$^{23}$ 
F.~Donovan,$^{14}$  
K.~L.~Dooley,$^{10}$  
S.~Doravari,$^{9}$  
I.~Dorrington,$^{95}$  
R.~Douglas,$^{43}$  
M.~Dovale~\'Alvarez,$^{53}$  
T.~P.~Downes,$^{24}$  
M.~Drago,$^{9}$  
R.~W.~P.~Drever$^{\sharp}$,$^{1}$
J.~C.~Driggers,$^{44}$  
Z.~Du,$^{79}$  
M.~Ducrot,$^{7}$ 
J.~Duncan,$^{94}$	
S.~E.~Dwyer,$^{44}$  
T.~B.~Edo,$^{100}$  
M.~C.~Edwards,$^{69}$  
A.~Effler,$^{6}$  
H.-B.~Eggenstein,$^{9}$  
P.~Ehrens,$^{1}$  
J.~Eichholz,$^{1}$  
S.~S.~Eikenberry,$^{5}$  
R.~A.~Eisenstein,$^{14}$	
R.~C.~Essick,$^{14}$  
Z.~B.~Etienne,$^{37,38}$  
T.~Etzel,$^{1}$  
M.~Evans,$^{14}$  
T.~M.~Evans,$^{6}$  
M.~Factourovich,$^{46}$  
V.~Fafone,$^{29,17,16}$ 
H.~Fair,$^{41}$  
S.~Fairhurst,$^{95}$  
X.~Fan,$^{79}$  
S.~Farinon,$^{55}$ 
B.~Farr,$^{85}$  
W.~M.~Farr,$^{53}$  
E.~J.~Fauchon-Jones,$^{95}$  
M.~Favata,$^{103}$  
M.~Fays,$^{95}$  
H.~Fehrmann,$^{9}$  
J.~Feicht,$^{1}$  
M.~M.~Fejer,$^{47}$ 
A.~Fernandez-Galiana,$^{14}$	
I.~Ferrante,$^{20,21}$ 
E.~C.~Ferreira,$^{15}$  
F.~Ferrini,$^{27}$ 
F.~Fidecaro,$^{20,21}$ 
I.~Fiori,$^{27}$ 
D.~Fiorucci,$^{35}$ 
R.~P.~Fisher,$^{41}$  
R.~Flaminio,$^{23,104}$ 
M.~Fletcher,$^{43}$  
H.~Fong,$^{105}$  
P.~W.~F.~Forsyth,$^{22}$  
S.~S.~Forsyth,$^{73}$  
J.-D.~Fournier,$^{62}$ 
S.~Frasca,$^{90,31}$ 
F.~Frasconi,$^{21}$ 
Z.~Frei,$^{51}$  
A.~Freise,$^{53}$  
R.~Frey,$^{67}$  
V.~Frey,$^{25}$ 
E.~M.~Fries,$^{1}$  
P.~Fritschel,$^{14}$  
V.~V.~Frolov,$^{6}$  
P.~Fulda,$^{5,76}$  
M.~Fyffe,$^{6}$  
H.~Gabbard,$^{9}$  
M.~Gabel,$^{106}$  
B.~U.~Gadre,$^{18}$  
S.~M.~Gaebel,$^{53}$  
J.~R.~Gair,$^{107}$  
L.~Gammaitoni,$^{39}$ 
M.~R.~Ganija,$^{78}$  
S.~G.~Gaonkar,$^{18}$  
F.~Garufi,$^{75,4}$ 
S.~Gaudio,$^{33}$  
G.~Gaur,$^{108}$  
V.~Gayathri,$^{109}$  
N.~Gehrels$^{\dag}$,$^{76}$  
G.~Gemme,$^{55}$ 
E.~Genin,$^{27}$ 
A.~Gennai,$^{21}$ 
D.~George,$^{11}$  
J.~George,$^{56}$  
L.~Gergely,$^{110}$  
V.~Germain,$^{7}$ 
S.~Ghonge,$^{73}$  
Abhirup~Ghosh,$^{19}$  
Archisman~Ghosh,$^{19,13}$  
S.~Ghosh,$^{61,13}$ 
J.~A.~Giaime,$^{2,6}$  
K.~D.~Giardina,$^{6}$  
A.~Giazotto,$^{21}$ 
K.~Gill,$^{33}$  
L.~Glover,$^{101}$  
E.~Goetz,$^{9}$  
R.~Goetz,$^{5}$  
S.~Gomes,$^{95}$  
G.~Gonz\'alez,$^{2}$  
J.~M.~Gonzalez~Castro,$^{20,21}$ 
A.~Gopakumar,$^{111}$  
M.~L.~Gorodetsky,$^{57}$  
S.~E.~Gossan,$^{1}$  
M.~Gosselin,$^{27}$ %
R.~Gouaty,$^{7}$ 
A.~Grado,$^{112,4}$ 
C.~Graef,$^{43}$  
M.~Granata,$^{23}$ 
A.~Grant,$^{43}$  
S.~Gras,$^{14}$  
C.~Gray,$^{44}$  
G.~Greco,$^{65,66}$ 
A.~C.~Green,$^{53}$  
P.~Groot,$^{61}$ 
H.~Grote,$^{9}$  
S.~Grunewald,$^{34}$  
P.~Gruning,$^{25}$ 
G.~M.~Guidi,$^{65,66}$ 
X.~Guo,$^{79}$  
A.~Gupta,$^{82}$  
M.~K.~Gupta,$^{99}$  
K.~E.~Gushwa,$^{1}$  
E.~K.~Gustafson,$^{1}$  
R.~Gustafson,$^{113}$  
B.~R.~Hall,$^{64}$  
E.~D.~Hall,$^{1}$  
G.~Hammond,$^{43}$  
M.~Haney,$^{111}$  
M.~M.~Hanke,$^{9}$  
J.~Hanks,$^{44}$  
C.~Hanna,$^{82}$  
O.~A.~Hannuksela,$^{87}$  
J.~Hanson,$^{6}$  
T.~Hardwick,$^{2}$  
J.~Harms,$^{65,66}$ 
G.~M.~Harry,$^{114}$  
I.~W.~Harry,$^{34}$  
M.~J.~Hart,$^{43}$  
C.-J.~Haster,$^{105}$  
K.~Haughian,$^{43}$  
J.~Healy,$^{115}$  
A.~Heidmann,$^{68}$ 
M.~C.~Heintze,$^{6}$  
H.~Heitmann,$^{62}$ 
P.~Hello,$^{25}$ 
G.~Hemming,$^{27}$ 
M.~Hendry,$^{43}$  
I.~S.~Heng,$^{43}$  
J.~Hennig,$^{43}$  
J.~Henry,$^{115}$  
A.~W.~Heptonstall,$^{1}$  
M.~Heurs,$^{9,32}$  
S.~Hild,$^{43}$  
D.~Hoak,$^{27}$ 
D.~Hofman,$^{23}$ 
K.~Holt,$^{6}$  
D.~E.~Holz,$^{85}$  
P.~Hopkins,$^{95}$  
C.~Horst,$^{24}$  
J.~Hough,$^{43}$  
E.~A.~Houston,$^{43}$  
E.~J.~Howell,$^{60}$  
Y.~M.~Hu,$^{9}$  
E.~A.~Huerta,$^{11}$  
D.~Huet,$^{25}$ 
B.~Hughey,$^{33}$  
S.~Husa,$^{96}$  
S.~H.~Huttner,$^{43}$  
T.~Huynh-Dinh,$^{6}$  
N.~Indik,$^{9}$  
D.~R.~Ingram,$^{44}$  
R.~Inta,$^{80}$  
G.~Intini,$^{90,31}$ 
H.~N.~Isa,$^{43}$  
J.-M.~Isac,$^{68}$ %
M.~Isi,$^{1}$  
B.~R.~Iyer,$^{19}$  
K.~Izumi,$^{44}$  
T.~Jacqmin,$^{68}$ 
K.~Jani,$^{73}$  
P.~Jaranowski,$^{116}$ 
S.~Jawahar,$^{117}$  
F.~Jim\'enez-Forteza,$^{96}$  
W.~W.~Johnson,$^{2}$  
D.~I.~Jones,$^{118}$  
R.~Jones,$^{43}$  
R.~J.~G.~Jonker,$^{13}$ 
L.~Ju,$^{60}$  
J.~Junker,$^{9}$  
C.~V.~Kalaghatgi,$^{95}$  
V.~Kalogera,$^{94}$  
S.~Kandhasamy,$^{6}$  
G.~Kang,$^{36}$  
J.~B.~Kanner,$^{1}$  
S.~Karki,$^{67}$  
K.~S.~Karvinen,$^{9}$	
M.~Kasprzack,$^{2}$  
M.~Katolik,$^{11}$  
E.~Katsavounidis,$^{14}$  
W.~Katzman,$^{6}$  
S.~Kaufer,$^{32}$  
K.~Kawabe,$^{44}$  
F.~K\'ef\'elian,$^{62}$ 
D.~Keitel,$^{43}$  
A.~J.~Kemball,$^{11}$  
R.~Kennedy,$^{100}$  
C.~Kent,$^{95}$  
J.~S.~Key,$^{119}$  
F.~Y.~Khalili,$^{57}$  
I.~Khan,$^{16,17}$ %
S.~Khan,$^{9}$  
Z.~Khan,$^{99}$  
E.~A.~Khazanov,$^{120}$  
N.~Kijbunchoo,$^{44}$  
Chunglee~Kim,$^{121}$  
J.~C.~Kim,$^{122}$  
W.~Kim,$^{78}$  
W.~S.~Kim,$^{123}$  
Y.-M.~Kim,$^{86,121}$  
S.~J.~Kimbrell,$^{73}$  
E.~J.~King,$^{78}$  
P.~J.~King,$^{44}$  
R.~Kirchhoff,$^{9}$  
J.~S.~Kissel,$^{44}$  
L.~Kleybolte,$^{30}$  
S.~Klimenko,$^{5}$  
P.~Koch,$^{9}$  
S.~M.~Koehlenbeck,$^{9}$  
S.~Koley,$^{13}$ %
V.~Kondrashov,$^{1}$  
A.~Kontos,$^{14}$  
M.~Korobko,$^{30}$  
W.~Z.~Korth,$^{1}$  
I.~Kowalska,$^{70}$ 
D.~B.~Kozak,$^{1}$  
C.~Kr\"amer,$^{9}$  
V.~Kringel,$^{9}$  
B.~Krishnan,$^{9}$  
A.~Kr\'olak,$^{124,125}$ 
G.~Kuehn,$^{9}$  
P.~Kumar,$^{105}$  
R.~Kumar,$^{99}$  
S.~Kumar,$^{19}$  
L.~Kuo,$^{83}$  
A.~Kutynia,$^{124}$ 
S.~Kwang,$^{24}$  
B.~D.~Lackey,$^{34}$  
K.~H.~Lai,$^{87}$  
M.~Landry,$^{44}$  
R.~N.~Lang,$^{24}$  
J.~Lange,$^{115}$  
B.~Lantz,$^{47}$  
R.~K.~Lanza,$^{14}$  
A.~Lartaux-Vollard,$^{25}$ 
P.~D.~Lasky,$^{126}$  
M.~Laxen,$^{6}$  
A.~Lazzarini,$^{1}$  
C.~Lazzaro,$^{50}$ 
P.~Leaci,$^{90,31}$ 
S.~Leavey,$^{43}$  
C.~H.~Lee,$^{86}$  
H.~K.~Lee,$^{127}$  
H.~M.~Lee,$^{121}$  
H.~W.~Lee,$^{122}$  
K.~Lee,$^{43}$  
J.~Lehmann,$^{9}$  
A.~Lenon,$^{37,38}$  
M.~Leonardi,$^{102,89}$ 
N.~Leroy,$^{25}$ 
N.~Letendre,$^{7}$ 
Y.~Levin,$^{126}$  
T.~G.~F.~Li,$^{87}$  
A.~Libson,$^{14}$  
T.~B.~Littenberg,$^{128}$  
J.~Liu,$^{60}$  
R.~K.~L.~Lo,$^{87}$ 
N.~A.~Lockerbie,$^{117}$  
L.~T.~London,$^{95}$  
J.~E.~Lord,$^{41}$  
M.~Lorenzini,$^{16,17}$ 
V.~Loriette,$^{129}$ 
M.~Lormand,$^{6}$  
G.~Losurdo,$^{21}$ 
J.~D.~Lough,$^{9,32}$  
C.~O.~Lousto,$^{115}$  
G.~Lovelace,$^{26}$  
H.~L\"uck,$^{32,9}$  
D.~Lumaca,$^{29,17}$ %
A.~P.~Lundgren,$^{9}$  
R.~Lynch,$^{14}$  
Y.~Ma,$^{59}$  
S.~Macfoy,$^{58}$  
B.~Machenschalk,$^{9}$  
M.~MacInnis,$^{14}$  
D.~M.~Macleod,$^{2}$  
I.~Maga\~na~Hernandez,$^{87}$  
F.~Maga\~na-Sandoval,$^{41}$  
L.~Maga\~na~Zertuche,$^{41}$  
R.~M.~Magee,$^{82}$ 
E.~Majorana,$^{31}$ 
I.~Maksimovic,$^{129}$ 
N.~Man,$^{62}$ 
V.~Mandic,$^{42}$  
V.~Mangano,$^{43}$  
G.~L.~Mansell,$^{22}$  
M.~Manske,$^{24}$  
M.~Mantovani,$^{27}$ 
F.~Marchesoni,$^{48,40}$ 
F.~Marion,$^{7}$ 
S.~M\'arka,$^{46}$  
Z.~M\'arka,$^{46}$  
C.~Markakis,$^{11}$  
A.~S.~Markosyan,$^{47}$  
E.~Maros,$^{1}$  
F.~Martelli,$^{65,66}$ 
L.~Martellini,$^{62}$ 
I.~W.~Martin,$^{43}$  
D.~V.~Martynov,$^{14}$  
K.~Mason,$^{14}$  
A.~Masserot,$^{7}$ 
T.~J.~Massinger,$^{1}$  
M.~Masso-Reid,$^{43}$  
S.~Mastrogiovanni,$^{90,31}$ 
A.~Matas,$^{42}$  
F.~Matichard,$^{14}$  
L.~Matone,$^{46}$  
N.~Mavalvala,$^{14}$  
N.~Mazumder,$^{64}$  
R.~McCarthy,$^{44}$  
D.~E.~McClelland,$^{22}$  
S.~McCormick,$^{6}$  
L.~McCuller,$^{14}$  
S.~C.~McGuire,$^{130}$  
G.~McIntyre,$^{1}$  
J.~McIver,$^{1}$  
D.~J.~McManus,$^{22}$  
T.~McRae,$^{22}$  
S.~T.~McWilliams,$^{37,38}$  
D.~Meacher,$^{82}$  
G.~D.~Meadors,$^{34,9}$  
J.~Meidam,$^{13}$ 
E.~Mejuto-Villa,$^{8}$  
A.~Melatos,$^{131}$  
G.~Mendell,$^{44}$  
R.~A.~Mercer,$^{24}$  
E.~L.~Merilh,$^{44}$  
M.~Merzougui,$^{62}$ 
S.~Meshkov,$^{1}$  
C.~Messenger,$^{43}$  
C.~Messick,$^{82}$  
R.~Metzdorff,$^{68}$ %
P.~M.~Meyers,$^{42}$  
F.~Mezzani,$^{31,90}$ %
H.~Miao,$^{53}$  
C.~Michel,$^{23}$ 
H.~Middleton,$^{53}$  
E.~E.~Mikhailov,$^{132}$  
L.~Milano,$^{75,4}$ 
A.~L.~Miller,$^{5}$  
A.~Miller,$^{90,31}$ 
B.~B.~Miller,$^{94}$  
J.~Miller,$^{14}$	
M.~Millhouse,$^{93}$  
O.~Minazzoli,$^{62}$ 
Y.~Minenkov,$^{17}$ 
J.~Ming,$^{34}$  
C.~Mishra,$^{133}$  
S.~Mitra,$^{18}$  
V.~P.~Mitrofanov,$^{57}$  
G.~Mitselmakher,$^{5}$ 
R.~Mittleman,$^{14}$  
A.~Moggi,$^{21}$ %
M.~Mohan,$^{27}$ 
S.~R.~P.~Mohapatra,$^{14}$  
M.~Montani,$^{65,66}$ 
B.~C.~Moore,$^{103}$  
C.~J.~Moore,$^{12}$  
D.~Moraru,$^{44}$  
G.~Moreno,$^{44}$  
S.~R.~Morriss,$^{97}$  
B.~Mours,$^{7}$ 
C.~M.~Mow-Lowry,$^{53}$  
G.~Mueller,$^{5}$  
A.~W.~Muir,$^{95}$  
Arunava~Mukherjee,$^{9}$  
D.~Mukherjee,$^{24}$  
S.~Mukherjee,$^{97}$  
N.~Mukund,$^{18}$  
A.~Mullavey,$^{6}$  
J.~Munch,$^{78}$  
E.~A.~M.~Muniz,$^{41}$  
P.~G.~Murray,$^{43}$  
K.~Napier,$^{73}$  
I.~Nardecchia,$^{29,17}$ 
L.~Naticchioni,$^{90,31}$ 
R.~K.~Nayak,$^{134}$	
G.~Nelemans,$^{61,13}$ 
T.~J.~N.~Nelson,$^{6}$  
M.~Neri,$^{54,55}$ 
M.~Nery,$^{9}$  
A.~Neunzert,$^{113}$  
J.~M.~Newport,$^{114}$  
G.~Newton$^{\ddag}$,$^{43}$  
K.~K.~Y.~Ng,$^{87}$  
T.~T.~Nguyen,$^{22}$  
D.~Nichols,$^{61}$ 
A.~B.~Nielsen,$^{9}$  
S.~Nissanke,$^{61,13}$ 
A.~Nitz,$^{9}$  
A.~Noack,$^{9}$  
F.~Nocera,$^{27}$ 
D.~Nolting,$^{6}$  
M.~E.~N.~Normandin,$^{97}$  
L.~K.~Nuttall,$^{41}$  
J.~Oberling,$^{44}$  
E.~Ochsner,$^{24}$  
E.~Oelker,$^{14}$  
G.~H.~Ogin,$^{106}$  
J.~J.~Oh,$^{123}$  
S.~H.~Oh,$^{123}$  
F.~Ohme,$^{9}$  
M.~Oliver,$^{96}$  
P.~Oppermann,$^{9}$  
Richard~J.~Oram,$^{6}$  
B.~O'Reilly,$^{6}$  
R.~Ormiston,$^{42}$  
L.~F.~Ortega,$^{5}$	
R.~O'Shaughnessy,$^{115}$  
D.~J.~Ottaway,$^{78}$  
H.~Overmier,$^{6}$  
B.~J.~Owen,$^{80}$  
A.~E.~Pace,$^{82}$  
J.~Page,$^{128}$  
M.~A.~Page,$^{60}$  
A.~Pai,$^{109}$  
S.~A.~Pai,$^{56}$  
J.~R.~Palamos,$^{67}$  
O.~Palashov,$^{120}$  
C.~Palomba,$^{31}$ 
A.~Pal-Singh,$^{30}$  
H.~Pan,$^{83}$  
B.~Pang,$^{59}$  
P.~T.~H.~Pang,$^{87}$  
C.~Pankow,$^{94}$  
F.~Pannarale,$^{95}$  
B.~C.~Pant,$^{56}$  
F.~Paoletti,$^{21}$ 
A.~Paoli,$^{27}$ 
M.~A.~Papa,$^{34,24,9}$  
H.~R.~Paris,$^{47}$  
W.~Parker,$^{6}$  
D.~Pascucci,$^{43}$  
A.~Pasqualetti,$^{27}$ 
R.~Passaquieti,$^{20,21}$ 
D.~Passuello,$^{21}$ 
B.~Patricelli,$^{135,21}$ 
B.~L.~Pearlstone,$^{43}$  
M.~Pedraza,$^{1}$  
R.~Pedurand,$^{23,136}$ 
L.~Pekowsky,$^{41}$  
A.~Pele,$^{6}$  
S.~Penn,$^{137}$  
C.~J.~Perez,$^{44}$  
A.~Perreca,$^{1,102,89}$ 
L.~M.~Perri,$^{94}$  
H.~P.~Pfeiffer,$^{105}$  
M.~Phelps,$^{43}$  
O.~J.~Piccinni,$^{90,31}$ 
M.~Pichot,$^{62}$ 
F.~Piergiovanni,$^{65,66}$ 
V.~Pierro,$^{8}$  
G.~Pillant,$^{27}$ 
L.~Pinard,$^{23}$ 
I.~M.~Pinto,$^{8}$  
M.~Pitkin,$^{43}$  
R.~Poggiani,$^{20,21}$ 
P.~Popolizio,$^{27}$ 
E.~K.~Porter,$^{35}$ 
A.~Post,$^{9}$  
J.~Powell,$^{43}$  
J.~Prasad,$^{18}$  
J.~W.~W.~Pratt,$^{33}$  
V.~Predoi,$^{95}$  
T.~Prestegard,$^{24}$  
M.~Prijatelj,$^{9}$  
M.~Principe,$^{8}$  
S.~Privitera,$^{34}$  
R.~Prix,$^{9}$  
G.~A.~Prodi,$^{102,89}$ 
L.~G.~Prokhorov,$^{57}$  
O.~Puncken,$^{9}$  
M.~Punturo,$^{40}$ 
P.~Puppo,$^{31}$ 
M.~P\"urrer,$^{34}$  
H.~Qi,$^{24}$  
J.~Qin,$^{60}$  
S.~Qiu,$^{126}$  
V.~Quetschke,$^{97}$  
E.~A.~Quintero,$^{1}$  
R.~Quitzow-James,$^{67}$  
F.~J.~Raab,$^{44}$  
D.~S.~Rabeling,$^{22}$  
H.~Radkins,$^{44}$  
P.~Raffai,$^{51}$  
S.~Raja,$^{56}$  
C.~Rajan,$^{56}$  
M.~Rakhmanov,$^{97}$  
K.~E.~Ramirez,$^{97}$ 
P.~Rapagnani,$^{90,31}$ 
V.~Raymond,$^{34}$  
M.~Razzano,$^{20,21}$ 
J.~Read,$^{26}$  
T.~Regimbau,$^{62}$ 
L.~Rei,$^{55}$ 
S.~Reid,$^{58}$  
D.~H.~Reitze,$^{1,5}$  
H.~Rew,$^{132}$  
S.~D.~Reyes,$^{41}$  
F.~Ricci,$^{90,31}$ 
P.~M.~Ricker,$^{11}$  
S.~Rieger,$^{9}$  
K.~Riles,$^{113}$  
M.~Rizzo,$^{115}$  
N.~A.~Robertson,$^{1,43}$  
R.~Robie,$^{43}$  
F.~Robinet,$^{25}$ 
A.~Rocchi,$^{17}$ 
L.~Rolland,$^{7}$ 
J.~G.~Rollins,$^{1}$  
V.~J.~Roma,$^{67}$  
R.~Romano,$^{3,4}$ 
C.~L.~Romel,$^{44}$  
J.~H.~Romie,$^{6}$  
D.~Rosi\'nska,$^{138,52}$ 
M.~P.~Ross,$^{139}$  
S.~Rowan,$^{43}$  
A.~R\"udiger,$^{9}$  
P.~Ruggi,$^{27}$ 
K.~Ryan,$^{44}$  
S.~Sachdev,$^{1}$  
T.~Sadecki,$^{44}$  
L.~Sadeghian,$^{24}$  
M.~Sakellariadou,$^{140}$  
L.~Salconi,$^{27}$ 
M.~Saleem,$^{109}$  
F.~Salemi,$^{9}$  
A.~Samajdar,$^{134}$  
L.~Sammut,$^{126}$  
L.~M.~Sampson,$^{94}$  
E.~J.~Sanchez,$^{1}$  
V.~Sandberg,$^{44}$  
B.~Sandeen,$^{94}$  
J.~R.~Sanders,$^{41}$  
B.~Sassolas,$^{23}$ 
B.~S.~Sathyaprakash,$^{82,95}$  
P.~R.~Saulson,$^{41}$  
O.~Sauter,$^{113}$  
R.~L.~Savage,$^{44}$  
A.~Sawadsky,$^{32}$  
P.~Schale,$^{67}$  
J.~Scheuer,$^{94}$  
E.~Schmidt,$^{33}$  
J.~Schmidt,$^{9}$  
P.~Schmidt,$^{1,61}$ 
R.~Schnabel,$^{30}$  
R.~M.~S.~Schofield,$^{67}$  
A.~Sch\"onbeck,$^{30}$  
E.~Schreiber,$^{9}$  
D.~Schuette,$^{9,32}$  
B.~W.~Schulte,$^{9}$  
B.~F.~Schutz,$^{95,9}$  
S.~G.~Schwalbe,$^{33}$  
J.~Scott,$^{43}$  
S.~M.~Scott,$^{22}$  
E.~Seidel,$^{11}$  
D.~Sellers,$^{6}$  
A.~S.~Sengupta,$^{141}$  
D.~Sentenac,$^{27}$ 
V.~Sequino,$^{29,17}$ 
A.~Sergeev,$^{120}$ 	
D.~A.~Shaddock,$^{22}$  
T.~J.~Shaffer,$^{44}$  
A.~A.~Shah,$^{128}$  
M.~S.~Shahriar,$^{94}$  
L.~Shao,$^{34}$  
B.~Shapiro,$^{47}$  
P.~Shawhan,$^{72}$  
A.~Sheperd,$^{24}$  
D.~H.~Shoemaker,$^{14}$  
D.~M.~Shoemaker,$^{73}$  
K.~Siellez,$^{73}$  
X.~Siemens,$^{24}$  
M.~Sieniawska,$^{52}$ 
D.~Sigg,$^{44}$  
A.~D.~Silva,$^{15}$  
A.~Singer,$^{1}$  
L.~P.~Singer,$^{76}$  
A.~Singh,$^{34,9,32}$  
R.~Singh,$^{2}$  
A.~Singhal,$^{16,31}$ 
A.~M.~Sintes,$^{96}$  
B.~J.~J.~Slagmolen,$^{22}$  
B.~Smith,$^{6}$  
J.~R.~Smith,$^{26}$  
R.~J.~E.~Smith,$^{1}$  
E.~J.~Son,$^{123}$  
J.~A.~Sonnenberg,$^{24}$  
B.~Sorazu,$^{43}$  
F.~Sorrentino,$^{55}$ 
T.~Souradeep,$^{18}$  
A.~P.~Spencer,$^{43}$  
A.~K.~Srivastava,$^{99}$  
A.~Staley,$^{46}$  
M.~Steinke,$^{9}$  
J.~Steinlechner,$^{43,30}$  
S.~Steinlechner,$^{30}$  
D.~Steinmeyer,$^{9,32}$  
B.~C.~Stephens,$^{24}$  
R.~Stone,$^{97}$  
K.~A.~Strain,$^{43}$  
G.~Stratta,$^{65,66}$ 
S.~E.~Strigin,$^{57}$  
R.~Sturani,$^{142}$  
A.~L.~Stuver,$^{6}$  
T.~Z.~Summerscales,$^{143}$  
L.~Sun,$^{131}$  
S.~Sunil,$^{99}$  
P.~J.~Sutton,$^{95}$  
B.~L.~Swinkels,$^{27}$ 
M.~J.~Szczepa\'nczyk,$^{33}$  
M.~Tacca,$^{35}$ 
D.~Talukder,$^{67}$  
D.~B.~Tanner,$^{5}$  
M.~T\'apai,$^{110}$  
A.~Taracchini,$^{34}$  
J.~A.~Taylor,$^{128}$  
R.~Taylor,$^{1}$  
T.~Theeg,$^{9}$  
E.~G.~Thomas,$^{53}$  
M.~Thomas,$^{6}$  
P.~Thomas,$^{44}$  
K.~A.~Thorne,$^{6}$  
K.~S.~Thorne,$^{59}$  
E.~Thrane,$^{126}$  
S.~Tiwari,$^{16,89}$ 
V.~Tiwari,$^{95}$  
K.~V.~Tokmakov,$^{117}$  
K.~Toland,$^{43}$  
M.~Tonelli,$^{20,21}$ 
Z.~Tornasi,$^{43}$  
C.~I.~Torrie,$^{1}$  
D.~T\"oyr\"a,$^{53}$  
F.~Travasso,$^{27,40}$ 
G.~Traylor,$^{6}$  
D.~Trifir\`o,$^{10}$  
J.~Trinastic,$^{5}$  
M.~C.~Tringali,$^{102,89}$ 
L.~Trozzo,$^{144,21}$ 
K.~W.~Tsang,$^{13}$ 
M.~Tse,$^{14}$  
R.~Tso,$^{1}$  
D.~Tuyenbayev,$^{97}$  
K.~Ueno,$^{24}$  
D.~Ugolini,$^{145}$  
C.~S.~Unnikrishnan,$^{111}$  
A.~L.~Urban,$^{1}$  
S.~A.~Usman,$^{95}$  
H.~Vahlbruch,$^{32}$  
G.~Vajente,$^{1}$  
G.~Valdes,$^{97}$	
M.~Vallisneri,$^{59}$
N.~van~Bakel,$^{13}$ 
M.~van~Beuzekom,$^{13}$ 
J.~F.~J.~van~den~Brand,$^{71,13}$ 
C.~Van~Den~Broeck,$^{13}$ 
D.~C.~Vander-Hyde,$^{41}$  
L.~van~der~Schaaf,$^{13}$ 
J.~V.~van~Heijningen,$^{13}$ 
A.~A.~van~Veggel,$^{43}$  
M.~Vardaro,$^{49,50}$ 
V.~Varma,$^{59}$  
S.~Vass,$^{1}$  
M.~Vas\'uth,$^{45}$ 
A.~Vecchio,$^{53}$  
G.~Vedovato,$^{50}$ 
J.~Veitch,$^{53}$  
P.~J.~Veitch,$^{78}$  
K.~Venkateswara,$^{139}$  
G.~Venugopalan,$^{1}$  
D.~Verkindt,$^{7}$ 
F.~Vetrano,$^{65,66}$ 
A.~Vicer\'e,$^{65,66}$ 
A.~D.~Viets,$^{24}$  
S.~Vinciguerra,$^{53}$  
D.~J.~Vine,$^{58}$  
J.-Y.~Vinet,$^{62}$ 
S.~Vitale,$^{14}$ 
T.~Vo,$^{41}$  
H.~Vocca,$^{39,40}$ 
C.~Vorvick,$^{44}$  
D.~V.~Voss,$^{5}$  
W.~D.~Vousden,$^{53}$  
S.~P.~Vyatchanin,$^{57}$  
A.~R.~Wade,$^{1}$  
L.~E.~Wade,$^{81}$  
M.~Wade,$^{81}$  
R.~Walet,$^{13}$ %
M.~Walker,$^{2}$  
L.~Wallace,$^{1}$  
S.~Walsh,$^{24}$  
G.~Wang,$^{16,66}$ 
H.~Wang,$^{53}$  
J.~Z.~Wang,$^{82}$  
M.~Wang,$^{53}$  
Y.-F.~Wang,$^{87}$  
Y.~Wang,$^{60}$  
R.~L.~Ward,$^{22}$  
J.~Warner,$^{44}$  
M.~Was,$^{7}$ 
J.~Watchi,$^{91}$  
B.~Weaver,$^{44}$  
L.-W.~Wei,$^{9,32}$  
M.~Weinert,$^{9}$  
A.~J.~Weinstein,$^{1}$  
R.~Weiss,$^{14}$  
L.~Wen,$^{60}$  
E.~K.~Wessel,$^{11}$  
P.~We{\ss}els,$^{9}$  
T.~Westphal,$^{9}$  
K.~Wette,$^{9}$  
J.~T.~Whelan,$^{115}$  
B.~F.~Whiting,$^{5}$  
C.~Whittle,$^{126}$  
D.~Williams,$^{43}$  
R.~D.~Williams,$^{1}$  
A.~R.~Williamson,$^{115}$  
J.~L.~Willis,$^{146}$  
B.~Willke,$^{32,9}$  
M.~H.~Wimmer,$^{9,32}$  
W.~Winkler,$^{9}$  
C.~C.~Wipf,$^{1}$  
H.~Wittel,$^{9,32}$  
G.~Woan,$^{43}$  
J.~Woehler,$^{9}$  
J.~Wofford,$^{115}$  
K.~W.~K.~Wong,$^{87}$  
J.~Worden,$^{44}$  
J.~L.~Wright,$^{43}$  
D.~S.~Wu,$^{9}$  
G.~Wu,$^{6}$  
W.~Yam,$^{14}$  
H.~Yamamoto,$^{1}$  
C.~C.~Yancey,$^{72}$  
M.~J.~Yap,$^{22}$  
Hang~Yu,$^{14}$  
Haocun~Yu,$^{14}$  
M.~Yvert,$^{7}$ 
A.~Zadro\.zny,$^{124}$ 
M.~Zanolin,$^{33}$  
T.~Zelenova,$^{27}$ 
J.-P.~Zendri,$^{50}$ 
M.~Zevin,$^{94}$  
L.~Zhang,$^{1}$  
M.~Zhang,$^{132}$  
T.~Zhang,$^{43}$  
Y.-H.~Zhang,$^{115}$  
C.~Zhao,$^{60}$  
M.~Zhou,$^{94}$  
Z.~Zhou,$^{94}$  
S.~J.~Zhu,$^{34,9}$	
X.~J.~Zhu,$^{60}$  
M.~E.~Zucker,$^{1,14}$  
and
J.~Zweizig$^{1}$%
\\
\medskip
(LIGO Scientific Collaboration and Virgo Collaboration) 
\\
\medskip
S.~Buchner$^{147,148}$,  
I.~Cognard$^{149,150}$,          
A.~Corongiu$^{151}$,             
P.~C.~C.~Freire$^{152}$,         
L.~Guillemot$^{149,150}$,        
G.~B.~Hobbs$^{153}$,             
M.~Kerr$^{154}$,                 
A.~G.~Lyne$^{155}$,              
A.~Possenti$^{151}$,             
A.~Ridolfi$^{152}$,              
R.~M.~Shannon$^{153,156}$,       
B.~W.~Stappers$^{155}$,          
and P.~Weltevrede$^{155}$        
\\
\medskip
{{}$^{*}$Deceased, March 2016. }%
{{}$^{\sharp}$Deceased, March 2017. }%
{${}^{\dag}$Deceased, February 2017. }%
{${}^{\ddag}$Deceased, December 2016. }%
}\noaffiliation
\affiliation {LIGO, California Institute of Technology, Pasadena, CA 91125, USA }
\affiliation {Louisiana State University, Baton Rouge, LA 70803, USA }
\affiliation {Universit\`a di Salerno, Fisciano, I-84084 Salerno, Italy }
\affiliation {INFN, Sezione di Napoli, Complesso Universitario di Monte S.Angelo, I-80126 Napoli, Italy }
\affiliation {University of Florida, Gainesville, FL 32611, USA }
\affiliation {LIGO Livingston Observatory, Livingston, LA 70754, USA }
\affiliation {Laboratoire d'Annecy-le-Vieux de Physique des Particules (LAPP), Universit\'e Savoie Mont Blanc, CNRS/IN2P3, F-74941 Annecy, France }
\affiliation {University of Sannio at Benevento, I-82100 Benevento, Italy and INFN, Sezione di Napoli, I-80100 Napoli, Italy }
\affiliation {Albert-Einstein-Institut, Max-Planck-Institut f\"ur Gravi\-ta\-tions\-physik, D-30167 Hannover, Germany }
\affiliation {The University of Mississippi, University, MS 38677, USA }
\affiliation {NCSA, University of Illinois at Urbana-Champaign, Urbana, IL 61801, USA }
\affiliation {University of Cambridge, Cambridge CB2 1TN, United Kingdom }
\affiliation {Nikhef, Science Park, 1098 XG Amsterdam, The Netherlands }
\affiliation {LIGO, Massachusetts Institute of Technology, Cambridge, MA 02139, USA }
\affiliation {Instituto Nacional de Pesquisas Espaciais, 12227-010 S\~{a}o Jos\'{e} dos Campos, S\~{a}o Paulo, Brazil }
\affiliation {Gran Sasso Science Institute (GSSI), I-67100 L'Aquila, Italy }
\affiliation {INFN, Sezione di Roma Tor Vergata, I-00133 Roma, Italy }
\affiliation {Inter-University Centre for Astronomy and Astrophysics, Pune 411007, India }
\affiliation {International Centre for Theoretical Sciences, Tata Institute of Fundamental Research, Bengaluru 560089, India }
\affiliation {Universit\`a di Pisa, I-56127 Pisa, Italy }
\affiliation {INFN, Sezione di Pisa, I-56127 Pisa, Italy }
\affiliation {OzGrav, Australian National University, Canberra, Australian Capital Territory 0200, Australia }
\affiliation {Laboratoire des Mat\'eriaux Avanc\'es (LMA), CNRS/IN2P3, F-69622 Villeurbanne, France }
\affiliation {University of Wisconsin-Milwaukee, Milwaukee, WI 53201, USA }
\affiliation {LAL, Univ. Paris-Sud, CNRS/IN2P3, Universit\'e Paris-Saclay, F-91898 Orsay, France }
\affiliation {California State University Fullerton, Fullerton, CA 92831, USA }
\affiliation {European Gravitational Observatory (EGO), I-56021 Cascina, Pisa, Italy }
\affiliation {Chennai Mathematical Institute, Chennai 603103, India }
\affiliation {Universit\`a di Roma Tor Vergata, I-00133 Roma, Italy }
\affiliation {Universit\"at Hamburg, D-22761 Hamburg, Germany }
\affiliation {INFN, Sezione di Roma, I-00185 Roma, Italy }
\affiliation {Leibniz Universit\"at Hannover, D-30167 Hannover, Germany }
\affiliation {Embry-Riddle Aeronautical University, Prescott, AZ 86301, USA }
\affiliation {Albert-Einstein-Institut, Max-Planck-Institut f\"ur Gravitations\-physik, D-14476 Potsdam-Golm, Germany }
\affiliation {APC, AstroParticule et Cosmologie, Universit\'e Paris Diderot, CNRS/IN2P3, CEA/Irfu, Observatoire de Paris, Sorbonne Paris Cit\'e, F-75205 Paris Cedex 13, France }
\affiliation {Korea Institute of Science and Technology Information, Daejeon 34141, Korea }
\affiliation {West Virginia University, Morgantown, WV 26506, USA }
\affiliation {Center for Gravitational Waves and Cosmology, West Virginia University, Morgantown, WV 26505, USA }
\affiliation {Universit\`a di Perugia, I-06123 Perugia, Italy }
\affiliation {INFN, Sezione di Perugia, I-06123 Perugia, Italy }
\affiliation {Syracuse University, Syracuse, NY 13244, USA }
\affiliation {University of Minnesota, Minneapolis, MN 55455, USA }
\affiliation {SUPA, University of Glasgow, Glasgow G12 8QQ, United Kingdom }
\affiliation {LIGO Hanford Observatory, Richland, WA 99352, USA }
\affiliation {Wigner RCP, RMKI, H-1121 Budapest, Konkoly Thege Mikl\'os \'ut 29-33, Hungary }
\affiliation {Columbia University, New York, NY 10027, USA }
\affiliation {Stanford University, Stanford, CA 94305, USA }
\affiliation {Universit\`a di Camerino, Dipartimento di Fisica, I-62032 Camerino, Italy }
\affiliation {Universit\`a di Padova, Dipartimento di Fisica e Astronomia, I-35131 Padova, Italy }
\affiliation {INFN, Sezione di Padova, I-35131 Padova, Italy }
\affiliation {MTA E\"otv\"os University, ``Lendulet'' Astrophysics Research Group, Budapest 1117, Hungary }
\affiliation {Nicolaus Copernicus Astronomical Center, Polish Academy of Sciences, 00-716, Warsaw, Poland }
\affiliation {University of Birmingham, Birmingham B15 2TT, United Kingdom }
\affiliation {Universit\`a degli Studi di Genova, I-16146 Genova, Italy }
\affiliation {INFN, Sezione di Genova, I-16146 Genova, Italy }
\affiliation {RRCAT, Indore MP 452013, India }
\affiliation {Faculty of Physics, Lomonosov Moscow State University, Moscow 119991, Russia }
\affiliation {SUPA, University of the West of Scotland, Paisley PA1 2BE, United Kingdom }
\affiliation {Caltech CaRT, Pasadena, CA 91125, USA }
\affiliation {OzGrav, University of Western Australia, Crawley, Western Australia 6009, Australia }
\affiliation {Department of Astrophysics/IMAPP, Radboud University Nijmegen, P.O. Box 9010, 6500 GL Nijmegen, The Netherlands }
\affiliation {Artemis, Universit\'e C\^ote d'Azur, Observatoire C\^ote d'Azur, CNRS, CS 34229, F-06304 Nice Cedex 4, France }
\affiliation {Institut de Physique de Rennes, CNRS, Universit\'e de Rennes 1, F-35042 Rennes, France }
\affiliation {Washington State University, Pullman, WA 99164, USA }
\affiliation {Universit\`a degli Studi di Urbino 'Carlo Bo', I-61029 Urbino, Italy }
\affiliation {INFN, Sezione di Firenze, I-50019 Sesto Fiorentino, Firenze, Italy }
\affiliation {University of Oregon, Eugene, OR 97403, USA }
\affiliation {Laboratoire Kastler Brossel, UPMC-Sorbonne Universit\'es, CNRS, ENS-PSL Research University, Coll\`ege de France, F-75005 Paris, France }
\affiliation {Carleton College, Northfield, MN 55057, USA }
\affiliation {Astronomical Observatory Warsaw University, 00-478 Warsaw, Poland }
\affiliation {VU University Amsterdam, 1081 HV Amsterdam, The Netherlands }
\affiliation {University of Maryland, College Park, MD 20742, USA }
\affiliation {Center for Relativistic Astrophysics and School of Physics, Georgia Institute of Technology, Atlanta, GA 30332, USA }
\affiliation {Universit\'e Claude Bernard Lyon 1, F-69622 Villeurbanne, France }
\affiliation {Universit\`a di Napoli 'Federico II', Complesso Universitario di Monte S.Angelo, I-80126 Napoli, Italy }
\affiliation {NASA Goddard Space Flight Center, Greenbelt, MD 20771, USA }
\affiliation {RESCEU, University of Tokyo, Tokyo, 113-0033, Japan. }
\affiliation {OzGrav, University of Adelaide, Adelaide, South Australia 5005, Australia }
\affiliation {Tsinghua University, Beijing 100084, China }
\affiliation {Texas Tech University, Lubbock, TX 79409, USA }
\affiliation {Kenyon College, Gambier, OH 43022, USA }
\affiliation {The Pennsylvania State University, University Park, PA 16802, USA }
\affiliation {National Tsing Hua University, Hsinchu City, 30013 Taiwan, Republic of China }
\affiliation {Charles Sturt University, Wagga Wagga, New South Wales 2678, Australia }
\affiliation {University of Chicago, Chicago, IL 60637, USA }
\affiliation {Pusan National University, Busan 46241, Korea }
\affiliation {The Chinese University of Hong Kong, Shatin, NT, Hong Kong }
\affiliation {INAF, Osservatorio Astronomico di Padova, Vicolo dell'Osservatorio 5, I-35122 Padova, Italy }
\affiliation {INFN, Trento Institute for Fundamental Physics and Applications, I-38123 Povo, Trento, Italy }
\affiliation {Universit\`a di Roma 'La Sapienza', I-00185 Roma, Italy }
\affiliation {Universit\'e Libre de Bruxelles, Brussels 1050, Belgium }
\affiliation {Sonoma State University, Rohnert Park, CA 94928, USA }
\affiliation {Montana State University, Bozeman, MT 59717, USA }
\affiliation {Center for Interdisciplinary Exploration \& Research in Astrophysics (CIERA), Northwestern University, Evanston, IL 60208, USA }
\affiliation {Cardiff University, Cardiff CF24 3AA, United Kingdom }
\affiliation {Universitat de les Illes Balears, IAC3---IEEC, E-07122 Palma de Mallorca, Spain }
\affiliation {The University of Texas Rio Grande Valley, Brownsville, TX 78520, USA }
\affiliation {Bellevue College, Bellevue, WA 98007, USA }
\affiliation {Institute for Plasma Research, Bhat, Gandhinagar 382428, India }
\affiliation {The University of Sheffield, Sheffield S10 2TN, United Kingdom }
\affiliation {California State University, Los Angeles, 5151 State University Dr, Los Angeles, CA 90032, USA }
\affiliation {Universit\`a di Trento, Dipartimento di Fisica, I-38123 Povo, Trento, Italy }
\affiliation {Montclair State University, Montclair, NJ 07043, USA }
\affiliation {National Astronomical Observatory of Japan, 2-21-1 Osawa, Mitaka, Tokyo 181-8588, Japan }
\affiliation {Canadian Institute for Theoretical Astrophysics, University of Toronto, Toronto, Ontario M5S 3H8, Canada }
\affiliation {Whitman College, 345 Boyer Avenue, Walla Walla, WA 99362 USA }
\affiliation {School of Mathematics, University of Edinburgh, Edinburgh EH9 3FD, United Kingdom }
\affiliation {University and Institute of Advanced Research, Gandhinagar Gujarat 382007, India }
\affiliation {IISER-TVM, CET Campus, Trivandrum Kerala 695016, India }
\affiliation {University of Szeged, D\'om t\'er 9, Szeged 6720, Hungary }
\affiliation {Tata Institute of Fundamental Research, Mumbai 400005, India }
\affiliation {INAF, Osservatorio Astronomico di Capodimonte, I-80131, Napoli, Italy }
\affiliation {University of Michigan, Ann Arbor, MI 48109, USA }
\affiliation {American University, Washington, D.C. 20016, USA }
\affiliation {Rochester Institute of Technology, Rochester, NY 14623, USA }
\affiliation {University of Bia{\l }ystok, 15-424 Bia{\l }ystok, Poland }
\affiliation {SUPA, University of Strathclyde, Glasgow G1 1XQ, United Kingdom }
\affiliation {University of Southampton, Southampton SO17 1BJ, United Kingdom }
\affiliation {University of Washington Bothell, 18115 Campus Way NE, Bothell, WA 98011, USA }
\affiliation {Institute of Applied Physics, Nizhny Novgorod, 603950, Russia }
\affiliation {Seoul National University, Seoul 08826, Korea }
\affiliation {Inje University Gimhae, South Gyeongsang 50834, Korea }
\affiliation {National Institute for Mathematical Sciences, Daejeon 34047, Korea }
\affiliation {NCBJ, 05-400 \'Swierk-Otwock, Poland }
\affiliation {Institute of Mathematics, Polish Academy of Sciences, 00656 Warsaw, Poland }
\affiliation {OzGrav, School of Physics \& Astronomy, Monash University, Clayton 3800, Victoria, Australia }
\affiliation {Hanyang University, Seoul 04763, Korea }
\affiliation {NASA Marshall Space Flight Center, Huntsville, AL 35811, USA }
\affiliation {ESPCI, CNRS, F-75005 Paris, France }
\affiliation {Southern University and A\&M College, Baton Rouge, LA 70813, USA }
\affiliation {OzGrav, University of Melbourne, Parkville, Victoria 3010, Australia }
\affiliation {College of William and Mary, Williamsburg, VA 23187, USA }
\affiliation {Indian Institute of Technology Madras, Chennai 600036, India }
\affiliation {IISER-Kolkata, Mohanpur, West Bengal 741252, India }
\affiliation {Scuola Normale Superiore, Piazza dei Cavalieri 7, I-56126 Pisa, Italy }
\affiliation {Universit\'e de Lyon, F-69361 Lyon, France }
\affiliation {Hobart and William Smith Colleges, Geneva, NY 14456, USA }
\affiliation {Janusz Gil Institute of Astronomy, University of Zielona G\'ora, 65-265 Zielona G\'ora, Poland }
\affiliation {University of Washington, Seattle, WA 98195, USA }
\affiliation {King's College London, University of London, London WC2R 2LS, United Kingdom }
\affiliation {Indian Institute of Technology, Gandhinagar Ahmedabad Gujarat 382424, India }
\affiliation {International Institute of Physics, Universidade Federal do Rio Grande do Norte, Natal RN 59078-970, Brazil }
\affiliation {Andrews University, Berrien Springs, MI 49104, USA }
\affiliation {Universit\`a di Siena, I-53100 Siena, Italy }
\affiliation {Trinity University, San Antonio, TX 78212, USA }
\affiliation {Abilene Christian University, Abilene, TX 79699, USA }

\affiliation{Square Kilometer Array South Africa, The Park, Park Road, Pinelands, Cape Town 7405, South 
Africa}
\affiliation{Hartebeesthoek Radio Astronomy Observatory, PO Box 443, Krugersdorp, 1740, South 
Africa}
\affiliation{Laboratoire de Physique et Chimie de l'Environnement et de l'Espace, LPC2E, 
CNRS-Universit\'{e} d'Orl\'{e}ans, F-45071 Orl\'{e}ans, France}
\affiliation{Station de Radioastronomie de Nan\c{c}ay, Observatoire de Paris, CNRS/INSU, F-18330 
Nan\c{c}ay, France}
\affiliation{INAF - Osservatorio Astronomico di Cagliari, via della Scienza 5, 09047 Selargius, Italy}
\affiliation{Max-Planck-Institut f\"{u}r Radioastronomie MPIfR, Auf dem H\"{u}gel 
69, D-53121 Bonn, Germany}
\affiliation{CSIRO Astronomy and Space Science, Australia Telescope National Facility, Box 76 
Epping, NSW, 1710, Australia}
\affiliation{Space Science Division, Naval Research Laboratory, Washington, DC 20375-5352, USA}
\affiliation{Jodrell Bank Centre for Astrophysics, School of Physics and Astronomy, University 
of Manchester, Manchester M13 9PL, UK}
\affiliation{International Centre for Radio Astronomy Research, Curtin University, Bentley, WA 
6101, Australia}

\date[\relax]{Dated: 9/22/2017}
\maketitle

 }{
 \author{The LIGO Scientific Collaboration and the Virgo Collaboration, \\
S.~Buchner,
I.~Cognard,
A.~Corongiu,
P.~C.~C.~Freire,
L.~Guillemot,
G.~B.~Hobbs,
M.~Kerr,
A.~G.~Lyne,
A.~Possenti,
A.~Ridolfi,
R.~M.~Shannon,
B.~W.~Stappers,
and P.~Weltevrede}
 }
}

\date{November 16, 2017}

\begin{abstract}
We present results from the first directed search for nontensorial
gravitational waves. While general relativity allows for tensorial (plus and
cross) modes only, a generic metric theory may, in principle, predict waves
with up to six different polarizations. This analysis is sensitive to
continuous signals of scalar, vector or tensor polarizations, and does not rely
on any specific theory of gravity. After searching data from the first
observation run of the advanced LIGO detectors for signals at twice the
rotational frequency of \npsrs~known pulsars, we find no evidence of
gravitational waves of any polarization. We report the first upper limits for
scalar and vector strains, finding values comparable in magnitude to
previously-published limits for tensor strain. Our results may be translated
into constraints on specific alternative theories of gravity.
\end{abstract}


\maketitle



\header{Introduction}. The first gravitational waves detected by the
Advanced Laser Interferometer Gravitational-wave Observatory (aLIGO) and Virgo
have already been used to place some of the most stringent constraints on
deviations from the general theory of relativity (GR) in the highly-dynamical
and strong-field regimes of gravity \cite{gw150914, gw151226, o1bbh, gw170104}.
However, even though some partial progress has been made with the observation
of GW170814 \cite{gw170814, Polarizations} and in spite of the wealth of new
information provided by GW170817 \cite{gw170817,grb170817}, it has not yet been
possible to unambiguously confirm GR's prediction that the associated metric
perturbations are of a tensor nature (helicity $\pm2$), rather than vector
(helicity $\pm1$), or scalar (helicity 0) \cite{gw150914_tgr}. This is
unfortunate, since the presence of nontensorial modes is a key prediction of
many extensions to GR \cite{Eardley1973a, Eardley1973b, tegp, Will2006,
Berti2015}. Most importantly, the detection of a scalar or vector component, no
matter how small, would automatically point to physics beyond Einstein's theory
\cite{tegp, Will2006}.

In order to experimentally study gravitational-wave polarizations directly, one needs a local
measurement of their geometric effect (i.e.~which directions are streched and
squeezed) that breaks degeneracies between the five distinguishable (to
differential-arm instruments) modes supported by a generic metric theory of
gravity \cite{Eardley1973a, Eardley1973b}.
For transient waves like those detected so far, this cannot be fully achieved
with the LIGO-Virgo network, as at least five noncooriented differential-arm
antennas are required to break {\em all} such degeneracies \cite{Will2006,
Chatziioannou2012}. Constraints on the magnitude of non-GR polarizations
inferred from indirect measurements, like the rate of orbital decay of binary
pulsars, are only meaningful in the context of specific theories (see e.g.\
\cite{Weisberg2010, Freire2012}, or \cite{Stairs2003, Wex2014} for reviews).

Theory-independent polarization measurements could instead be carried out with
current detectors in the presence of signals sufficiently long to probe the
detector antenna patterns, which are themselves polarization-sensitive
\cite{Isi2015, Isi2017, Nishizawa2009, Callister2017}. Such is the case, for
instance, for the continuous, almost-monochromatic waves expected from spinning
neutron stars with an asymmetric moment of inertia \cite{Thorne1987}. Known
galactic pulsars are one of the main candidates for searches for such signals
in data from ground-based detectors, and analyses targeting them have already
achieved sensitivities that are comparable to, or even surpass, their canonical
spin-down limit (i.e.~the strain that would be produced if the observed
slowdown in the pulsar's rotation was completely due to gravitational
radiation) \cite{o1cw}. 

However, all previous targeted searches have been, by design, restricted to
tensorial gravitational polarizations {\em only}. This leaves open the
possibility that, due to a departure from GR, the neutron stars targeted in
previous searches may indeed be emitting strong continuous waves with
nontensorial content, in spite of the null results of standard searches.

In this paper, we present results from a search for continuous gravitational waves
in aLIGO data that makes no assumptions about how the 
gravitational field transforms under local spatial rotations, and is thus sensitive
to any of the five measurable polarizations allowed by a generic metric theory of gravity.
We targeted \npsrs\ known pulsars using data from aLIGO's first observation run
(O1), and assuming emission at twice the rotational frequency of the source.

Our data provide no evidence for the emission of gravitational signals of
tensorial or nontensorial polarization from any of the pulsars targeted. For
sources in the most sensitive band of our detectors, we constrain the strain of
the scalar and vector modes to be below \red{$1.5\times10^{-26}$} at 95\%
credibility. These are the first direct upper limits for scalar and vector
strain, and may in principle be used to constrain beyond-GR
theories of gravity.

\header{Analysis.} We search aLIGO O1 data from the Hanford (H1) and
Livingston (L1) detectors for continuous waves of any polarization (tensor,
scalar or vector) by applying the Bayesian time-domain method of
\cite{Dupuis2005}, generalized to non-GR modes as described in \cite{Isi2017}
and summarized below. Our analysis follows closely that of \cite{o1cw}, and
uses the exact same interferometric data.

Calibrated detector data are heterodyned and filtered using the timing
solutions obtained from electromagnetic observations for each pulsar. The
maximum calibration uncertainties estimated over the whole run give a limit on
the combined H1 and L1 amplitude uncertainties of \red{14}\%---this is the
conservative level of uncertainty on the strain upper limits
\cite{O1cal2016,o1cw}.

The data streams start on 2015 Sep 11 at 01:25:03 UTC for H1 and 18:29:03 UTC
for L1, and finish on 2016 Jan 19 at 17:07:59 UTC at both sites. The pulsar
timing solutions used are also the same as in \cite{o1cw} and were obtained
from the 42-ft telescope and Lovell telescope at Jodrell Bank (UK), the 26-m
telescope at Hartebeesthoek (South Africa), the Parkes radio telescope
(Australia), the Nancay Decimetric Radio Telescope (France), the Arecibo
Observatory (Puerto Rico) and the Fermi Large Area Telescope (LAT).

As described in detail in \cite{Isi2017}, we construct a Bayesian hypothesis
that captures signals of any polarization content (our {\em any-signal}
hypothesis, $\hyps$) by combining the sub-hypotheses corresponding to the
signal being composed of tensor, vector, scalar modes, or any combination
thereof. Each of these sub-hypotheses corresponds to a different signal model;
in particular, the least restrictive template includes contributions from all
polarizations and can be written as:
\beq \label{eq:raw_signal}
h(t) = \sum_p F_p(t; \alpha, \delta, \psi) h_p(t),
\eeq
where the sum is over the five independent polarizations: plus ($+$), cross
($\times$), vector-x (x), vector-y (y) and scalar (s) \cite{Eardley1973b}. The
two scalar modes in the most common basis, breathing and longitudinal,
are degenerate for networks of quadrupolar antennas \cite{Will2006}, so we do
not make a distinction between them.

Each term in \eq{raw_signal} is the product of an antenna pattern function
$F_p$ and an intrinsic strain function $h_p$. We define the different
polarizations in a wave-frame such that the $z$-axis points in the direction of
propagation, $x$ lies in the plane of the sky along the line of nodes (here
defined to be the intersection of the equatorial plane of the source with the
plane of the sky), and $y$ completes the right-handed system, such that the
polarization angle $\psi$ is the angle between the $y$-axis and the projection
of the celestial North onto the plane of the sky (see e.g.\
\cite{Anderson2001}). We can thus write the $F_p$'s as implicit functions of
the source's right ascension $\alpha$, declination $\delta$ and polarization
$\psi$. (For the sources targeted here, $\alpha$ and $\delta$ are always known
to high accuracy, while $\psi$ is usually unknown.) The antenna patterns
acquire their time dependence from the sidereal rotation of the Earth; explicit
expressions for the $F_p$'s are given in \cite{Isi2017, Isi2015, Nishizawa2009,
Blaut2012, poisson2014gravity}.

For a continuous wave, the polarizations take the simple form:
\beq \label{eq:signal}
h_p(t) = a_p \cos(\phi(t)+\phi_p),
\eeq
where $a_p$ is a time-independent strain amplitude, $\phi(t)$ is the intrinsic
phase evolution, and $\phi_p$ a phase offset for each polarization. The nature
of these three quantities depends on the specifics of the underlying theory of
gravity and the associated emission mechanism (for different emission
mechanisms within GR, see e.g.\ \cite{Zimmermann1979, Owen1998,
Bondarescu2009}). While we treat $a_p$ and $\phi_p$ as free parameters, we take
$\phi(t)$ to be the same as in the traditional GR analysis \cite{o1cw}:
\beq
\phi(t) = 2\pi \sum^N_{j=0} \frac{\partial_t^{(j)} f_{\rm GW,0}}{(j+1)!} \left[t - T_0 +
\delta t(t) \right]^{(j+1)},
\eeq
where $\partial_t^{(j)} f_{\rm GW,0}$ is the $j$\ts{th} time derivative of $f_{\rm GW,0}$, the
emission frequency measured at the fiducial time $T_0$; $\delta t
(t)$ is the time delay from the observatory to the solar system barycenter
(including the known R\o mer, Shapiro and Einstein delays), and can also
include binary system corrections to transform the time coordinate to a
frame approximately inertial with respect to the source; $N$ is the order of
the series expansion (1 or 2 for most sources).

The gravitational-wave frequency $\fgw$ is related to the rotational frequency
of the source $\frot$, which is in turn known from electromagnetic
observations. Although arbitrary theories of gravity and emission mechanisms
may predict gravitational emission at any multiple of the rotational frequency,
here we assume $\fgw=2\frot$, in accordance with the most favored emission
model in GR \cite{Thorne1987}. This restriction arises from practical
considerations affecting our specific implementation, and will be relaxd in 
future studies.

For convenience, we define {\em effective strain amplitudes} for tensor, vector
and scalar modes respectively by
\beq \label{eq:ht}
\hT \equiv \sqrt{a_+^2 + a_\times^2},
\eeq
\beq \label{eq:hv}
\hV \equiv \sqrt{a_{\rm x}^2 + a_{\rm y}^2},
\eeq
\beq \label{eq:hs}
\hS \equiv a_{\rm s},
\eeq
in terms of the intrinsic $a_p$ amplitudes of \eq{signal}. These quantities may
serve as proxy for the total power in each polarization group.

\begin{table}
\caption{Existing orientation information for pulsars in our band, obtained
from observations of the pulsar wind nebulae (see Table 3 in \cite{tcw2013},
and \cite{Ng2004, Ng2008} for measurement details).}
\label{tab:orientation}
\begin{ruledtabular}
\begin{tabular}{ l c c }
 & $\iota$ & $\psi$ \\
\midrule
J0534+2200 & $62\degree.2 \pm 1\degree.9$ & $35\degree.2 \pm 1\degree.5$\\
J0537--6910 & $92\degree.8 \pm 0\degree.9$ & $41\degree.0 \pm 2\degree.2$\\
J0835--4510 & $63\degree.6 \pm 0\degree.6$ & $40\degree.6 \pm 0\degree.1$\\
J1833--1034 & $85\degree.4 \pm 0\degree.3$ & $45\degree \pm 1\degree $\\
J1952+3252 & N/A & $-11\degree.5 \pm 8\degree.6$\\
\end{tabular}
\end{ruledtabular}
\end{table}

One may recover the GR hypothesis considered in previous analyses by setting:
\beq \label{eq:a_plus}
a_+ = h_0 (1 + \cos^2 \iota)/2~,~\phi_+ = \phi_0,
\eeq
\beq \label{eq:a_cross}
a_\times = h_0 \cos\iota~,~\phi_\times = \phi_0-\pi/2,
\eeq
\beq
a_{\rm x} = a_{\rm y} = a_{\rm s} = 0,
\eeq
where $\iota$ is the inclination (angle between the line of sight and the spin
axis of the source), and $h_0$, $\phi_0$ are free parameters. (As with $\psi$,
$\iota$ is unknown for most pulsars.) This corresponds to the standard
triaxial-star emission mechanism (see e.g. \cite{Jones2002}). We use this
parameterization only when we wish to incorporate known orientation information
as explained below; otherwise, we parametrize the tensor polarizations directly
in terms of $a_+$, $a_\times$, $\phi_+$ and $\phi_\times$.

Templates of the form of \eq{raw_signal}, together with appropriate priors,
allow us to compute Bayes factors (marginalized-likelihood ratios) for the
presence of signals in the data vs Gaussian noise. We do this using an
extension of the nested sampling implementation presented in \cite{Pitkin2012}
(see \cite{Isi2017} for details specific to non-GR polarizations). The
Bayes factors corresponding to each signal model may be combined into the odds
$\oddssn$ that the data contain a continuous signal of any polarization vs
Gaussian noise:
\beq \label{eq:oddssn}
\oddssn = \p(\hyps\mid\data) / \p(\hypn\mid\data)\, ,
\eeq
i.e.\ the ratio of the posterior probabilities that the data $\data$ contain a
signal of any polarizations ($\hyps$) vs just Gaussian noise ($\hypn$). We
compute these odds by setting model priors such that $\p(\hyps) = \p(\hypn)$;
then, by Bayes' theorem, $\oddssn = \bayessn$, with the Bayes factor
\beq
\bayessn \equiv \p(\data\mid\hyps) / \p(\data\mid\hypn)\, .
\eeq

Built into the astrophysical signal hypothesis, $\hyps$, is the
requirement of coherence across detectors, which must be satisfied by
a real gravitational wave. In order to make the analysis more robust against
non-Gaussian instrumental features in the data, we also define an {\em
instrumental feature} hypothesis, $\hyp{I}$, that identifies non-Gaussian noise
artifacts by their lack of coherence across detectors \cite{o1cw, Keitel2014}.
In particular, we define $\hyp{I}$ to capture Gaussian noise {\em or} a
detector-incoherent signal (i.e.\ a feature that mimics an astrophysical signal
in a single instrument, but is not recovered consistently across the network)
in each detector \cite{Isi2017}. We may then compare this to $\hyps$ by means
of the odds $\oddsci$. For $D$ detectors, this is given by:
\beq \label{eq:bci}
\log \oddsci = \log \bayessn - \sum^\ndet_{d=1} \log\left({\cal B}^{{\rm
S}_d}_{{\rm N}_d} + 1\right),
\eeq
where ${\cal B}^{{\rm S}_d}_{{\rm N}_d}$ is the signal vs noise Bayes factor
computed only from data from the $d$\ts{th} detector. This choice implicitly
assigns prior weight to the models such that $\p(\hyps)=\p(\hyp{I}) \times
0.5^{\ndet}$ \cite{Isi2017}. For an in depth analysis of the behavior of the
different Bayesian hypotheses considered here, in the presence and absence of
simulated signals of all polarizations, we again refer the reader to the methods 
paper \cite{Isi2017}.

\begin{figure}
\centering
\includegraphics[width=\columnwidth]{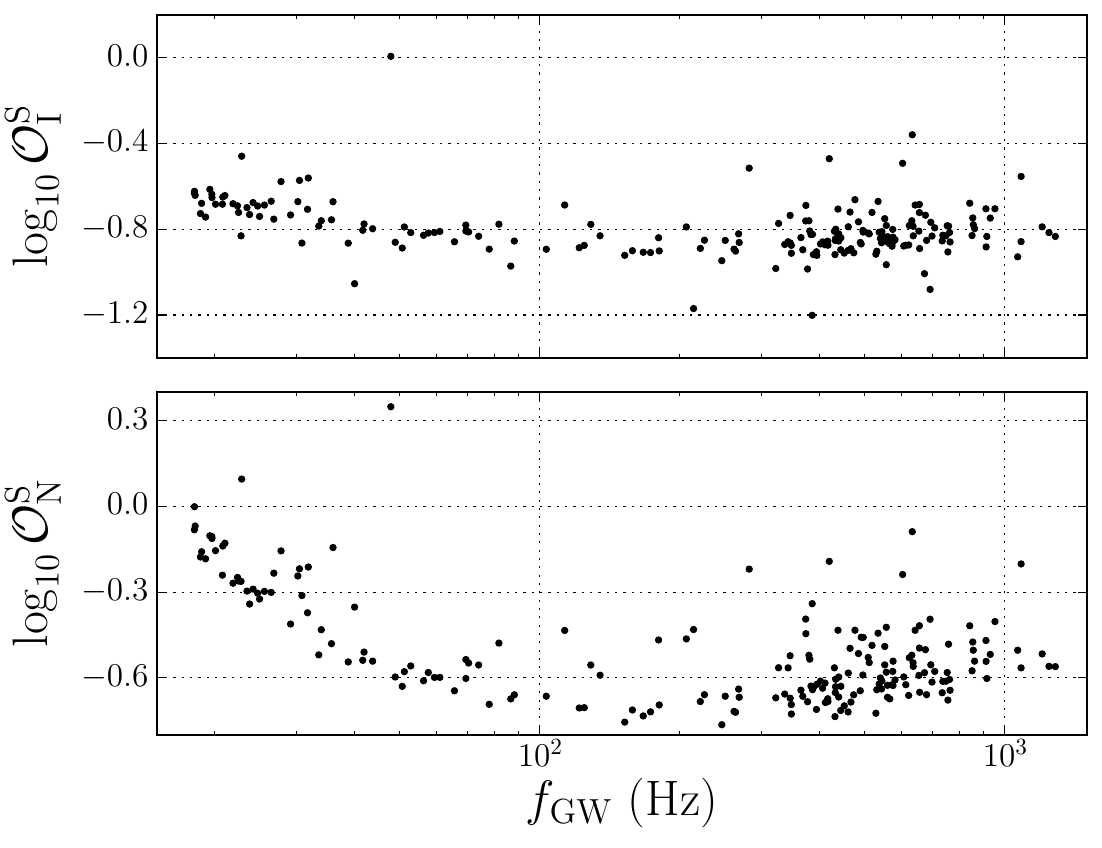}
\caption{{\em Log-odds vs emission frequency}. Log-odds comparing the any-signal
hypothesis to the instrumental (top) and Gaussian noise (bottom) hypotheses, as
a function of assumed gravitational-wave frequency, $\fgw=2\frot$, for each pulsar. Looking at
the top plot for $\log_{10}\oddsci$, notice that the instrumental noise hypothesis
is clearly favored for all pulsars except one, for which the analysis is
inconclusive. (This is J1932+17, the same non-significant outlier identified in
\cite{o1cw}.) These results were obtained without incorporating any information
on the source orientation, and are tabulated in Table II in the supplementary
material \cite{suppmat}. Expressions for both odds are given in \eq{oddssn} and
\eq{bci}.}
\label{fig:odds_s_n_freq}
\end{figure}

\begin{figure}
\centering
\includegraphics[width=\columnwidth]{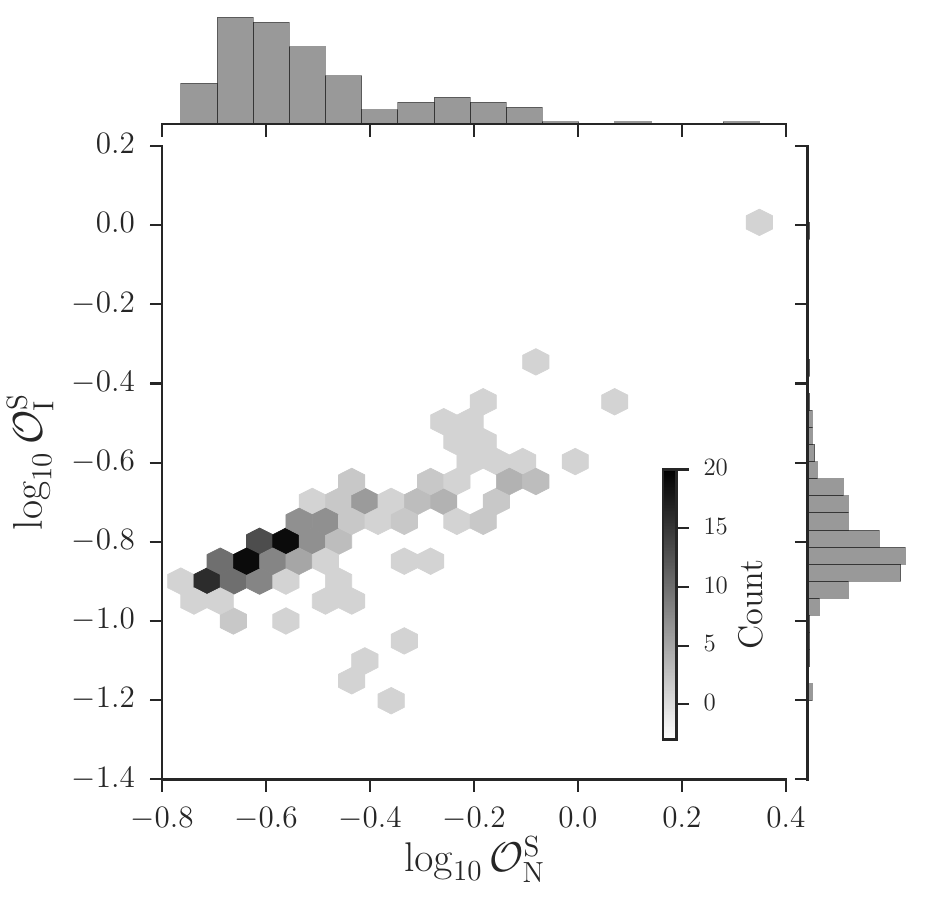}
\caption{{\em Log-odds distributions}. Distributions of log-odds comparing the
any-signal hypothesis to the instrumental (ordinate axis, right) and Gaussian
noise (abscissa axis, top) hypotheses for all pulsars. This plot contains the
same information as \fig{odds_s_n_freq} and displays the same non-significant
outlier. These results were obtained without incorporating any information on
the source orientation, and are tabulated in Table II in the supplementary
material \cite{suppmat}. Expressions for both odds in this plot are given in
\eq{oddssn} and \eq{bci}. We underscore that, although this plot looks similar
to Fig.\ 2 in \cite{o1cw}, the signal hypothesis here incorporates scalar,
vector and tensor modes, in all possible combinations.}
\label{fig:odds_c_i_hist}
\end{figure}

We compute likelihoods by taking source location, frequency and frequency
derivatives as known quantities. In computing Bayes factors, we employ priors
uniform in the logarithm of amplitude parameters ($h_0$ or $a_p$'s), since
these are the least informative priors for scaling coefficients
\cite{Jaynes1968}; we bound these amplitudes to the $10^{-28}$--$10^{-24}$
range \footnote{The specific range chosen for the amplitude priors has little
effect on our results, as explained in Appendix B of \cite{Isi2017}}. On the
other hand, flat amplitude priors are used to compute upper limits, to
facilitate comparison with published GR results in \cite{o1cw}. In all cases,
flat priors are placed over all phase offsets ($\phi_0$ and all the
$\phi_p$'s). 

For those few cases in which some orientation information exists (see Table
\ref{tab:orientation}), we analyze the data a second time using the triaxial
parametrization of tensor modes, Eqs.\ \eqref{eq:a_plus} and
\eqref{eq:a_cross}, taking that information into account by marginalizing over
ranges of $\cos\iota$ and $\psi$ in agreement with measurement uncertainties.
Following previous work \cite{o1cw}, we only consider orientation constraints obtained
from pulsar wind nebulae. However, pulsar orientations can also be inferred
from other measurements, especially if the object is in a binary (e.g.
\cite{Ferdman2013, Rickett2014, Zhu2015}). We will consider incorporating such
constraints in future searches.

\begin{figure*}
\centering
\includegraphics[width=\textwidth]{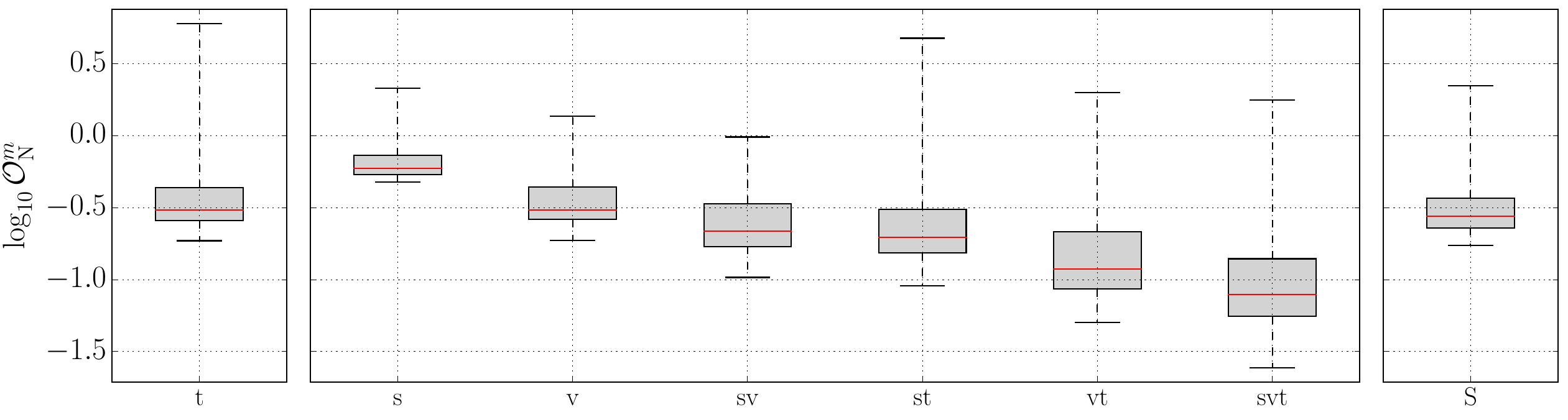}
\caption{{\em Sub-hypothesis odds}. Box plots for the distribution of the
signal vs noise log-odds for each of the sub-hypotheses considered, for all of
the pulsars analyzed. The sub-hypotheses are: (st), vector-tensor (tv),
scalar-vector-tensor tensor-only (t), scalar-only (s), vector-only (v),
scalar-vector (sv), scalar-tensor (st), vector-tensor (vt), and
scalar-vector-tensor (stv); these are all combined into the signal hypothesis
(S). The quantity represented is $\log_{10}{\cal B}^m_{\rm N}$, which is the same as
$\log_{10}{\cal O}^m_{\rm N}$ if neither ${\cal H}_m$ nor $\hyp{N}$ are favored {\em
a priori} (hence the label on the ordinate axis). The horizontal red line marks
the median of the distribution, while each gray box extends from the lower to
upper quartile, and the whiskers mark the full range of the distribution of
$\log_{10}{\cal O}^m_{\rm N}$ for the \npsrs\ pulsars analyzed. These results were
produced without incorporating any information on the source orientation, and
are tabulated in Table II in the supplementary material \cite{suppmat}.}
\label{fig:box} \end{figure*}

\header{Results.} We find no evidence of continuous-wave signals of any
polarization, tensorial or otherwise, from any of the \npsrs\ pulsars analyzed.
The main quantity of interest is $\log_{10}\oddsci$, defined in \eq{bci}, since it
encodes the probability that the data contain a signal vs just instrumental
noise (Gaussian or otherwise). This quantity, together with the log-odds for
signal vs Gaussian noise, is presented as a function of assumed emission frequency
for each pulsar in \fig{odds_s_n_freq}, and histogrammed in
\fig{odds_c_i_hist}. Importantly, the outliers in \fig{odds_s_n_freq} lose
significance once $\log_{10}\oddsci$ is taken into account; indeed, Figs.\
\ref{fig:odds_s_n_freq} and \ref{fig:odds_c_i_hist} reveal the usefulness of
$\log_{10}\oddsci$ in increasing the robustness of the search against non-Gaussian
instrumental artifacts.

Based on the intrinsic probabilistic meaning of $\log_{10}\oddsci$ in terms of
betting odds, it is standard to demand at least $\log_{10}\oddsci > 1$ to conclude
that the signal model is favored (see e.g.\ the table in Sec.\ 3.2 of
\cite{Kass1995}, or Jeffrey's original criteria in \cite{Jeffreys1998} or
\cite{Robert2009}). Since none of the odds obtained meet this criterion, we
conclude that there is no evidence for signals from any of the pulsars
targeted. In most cases, $\log_{10}\oddsci < 0$ and the noise model is clearly
favored; the single exception is J1932+17, for which $\log_{10}\oddsci \sim 0$, so
that we can make no conclusive statement about which hypothesis is preferred.
(The presence of this non-significant outlier is to be expected, as it was
already identified in \cite{o1cw}.)

The distribution of the odds corresponding to the subhypotheses making up
$\hyps$ are summarized in the box plots of \fig{box}. These correspond to
tensor-only (t), scalar-only (s), vector-only (v), scalar-vector (sv),
scalar-tensor (st), vector-tensor (vt), and scalar-vector-tensor (stv) models.
The mean of these distributions decreases with the number of degrees of freedom
in the model, which is to be expected from the associated Occam penalties
\cite{Isi2017}. The right-most panel in \fig{box} shows the distribution of
$\log_{10}\oddssn$, which results from the combination of all the other odds; 
this is the same quantity histogrammed on the abscissa of \fig{odds_c_i_hist}.

In the absence of any discernible signals, we produce upper limits for the
magnitude of scalar, vector and tensor polarizations, with a 95\% credibility.
As usual in Bayesian analyses, upper limits are obtained by integrating
posterior probability distributions for the relevant parameters up to the
desired credibility (see e.g.\ \cite{Isi2017}). Using the effective amplitude
definitions of Eqs.\ \eqref{eq:ht}--\eqref{eq:hs}, these quantities are
presented in \fig{ul} as a function of assumed emission frequency, and the
supplementary material. The plotted upper limits are computed under the
assumption of a signal model that includes all five independent polarizations
($\hyp{svt}$); \blue{limits obtained assuming other signal models may be found
online in \cite{suppmat}}. Previous work has demonstrated that the presence or
absence of a GR component does not affect the non-GR upper limits (Fig.\
\red{13} in \cite{Isi2017}). 

As expected, the upper limits presented here are comparable in magnitude to the
upper limits on the GR strain obtained by the traditional searches \cite{o1cw}.
However, constraints on the scalar amplitude are, on average, around 20\% less
stringent than those on the vector or tensor amplitudes. This is a consequence
of the fact that, for most source locations in the sky, the LIGO detectors are
intrinsically less sensitive to continuous waves of scalar (breathing or
longitudinal) polarization \cite{Isi2017}.

Technically, traditional all-sky searches for continuous gravitational waves
are also sensitive to nontensorial modes, because they are generally designed
to look for any signal of sidereal and half-sidereal periodicities in the data,
without assuming knowledge of phase evolution or source sky-location
\cite{einsthome2016, cweh2017, cwallsky2016, cwallsky2017, cwallskybin2014}.
However, as can be seen by comparing the magnitude of all-sky upper limits
(e.g.\ Fig.\ 9 in \cite{einsthome2016}) to those in shown here in \fig{ul}, the
sensitivity of these searches would be substantially poorer than that of a
targeted search like this one---if only because they are not targeted to a
specific source. This is especially true if the search is optimized for a given
signal polarization (e.g.\ circular combination of plus and cross).

Odds and 95\%-credible upper limits are summarized in the supplementary
material: Table I, for pulsars with measured orientations (using the triaxial
parameterization of tensor modes), and Table II, for all pulsars without
incorporating any orientation information (using the unconstrained
parameterization of tensor modes) \cite{suppmat}. Odds values are reported with
an error of \red{5}\% at \red{90}\% confidence; errors on the upper limits due
to the use of finite samples in estimating posterior probability distributions
are at most \red{10}\% at \red{90}\% confidence, which is slightly less than
the \red{15}\% error expected from calibration uncertainties.

\begin{figure*}%
\centering
\subfloat{
\includegraphics[width=\textwidth]{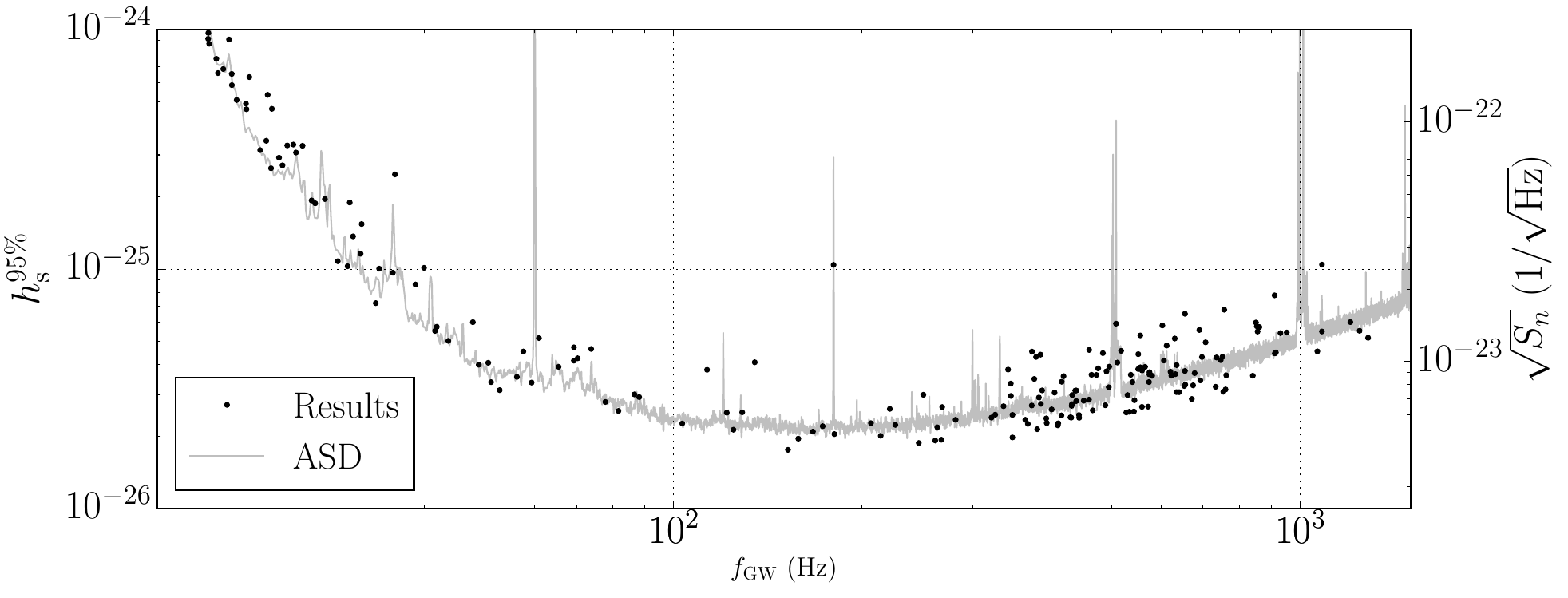}}\\
\subfloat{
\centering
\includegraphics[width=\textwidth]{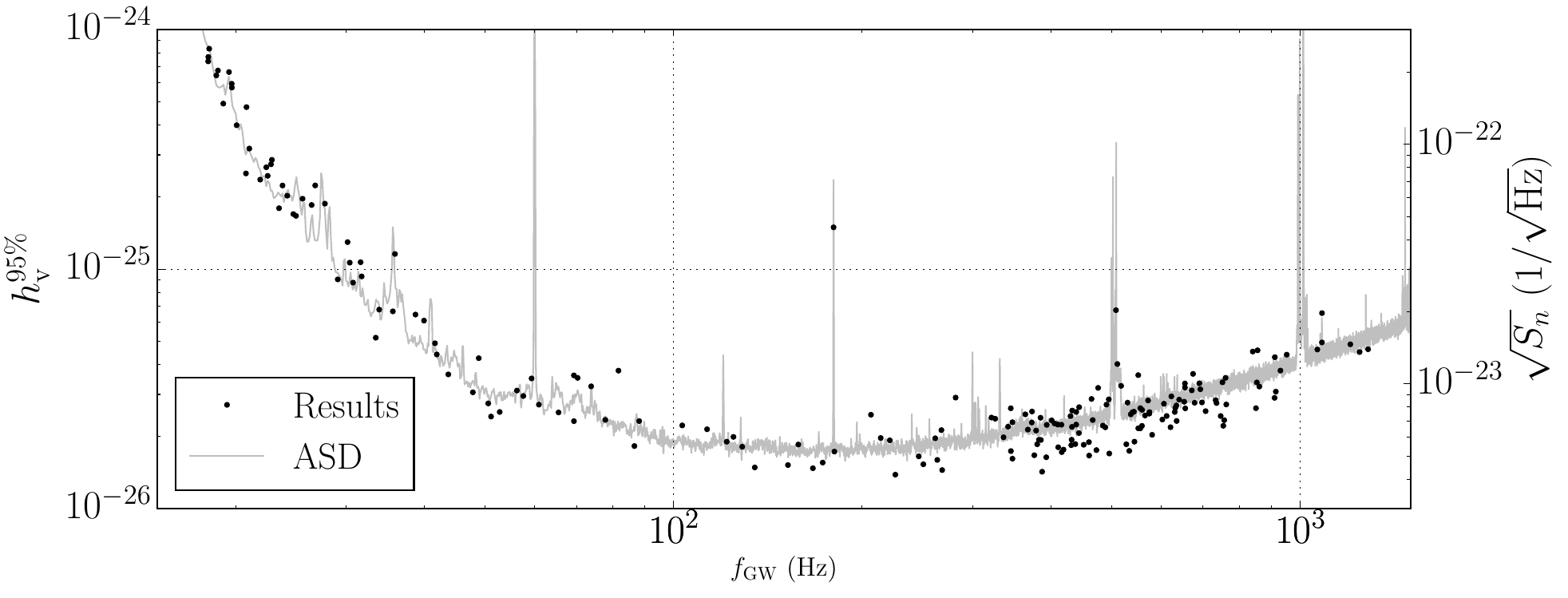}}\\
\centering
\subfloat{
\includegraphics[width=\textwidth]{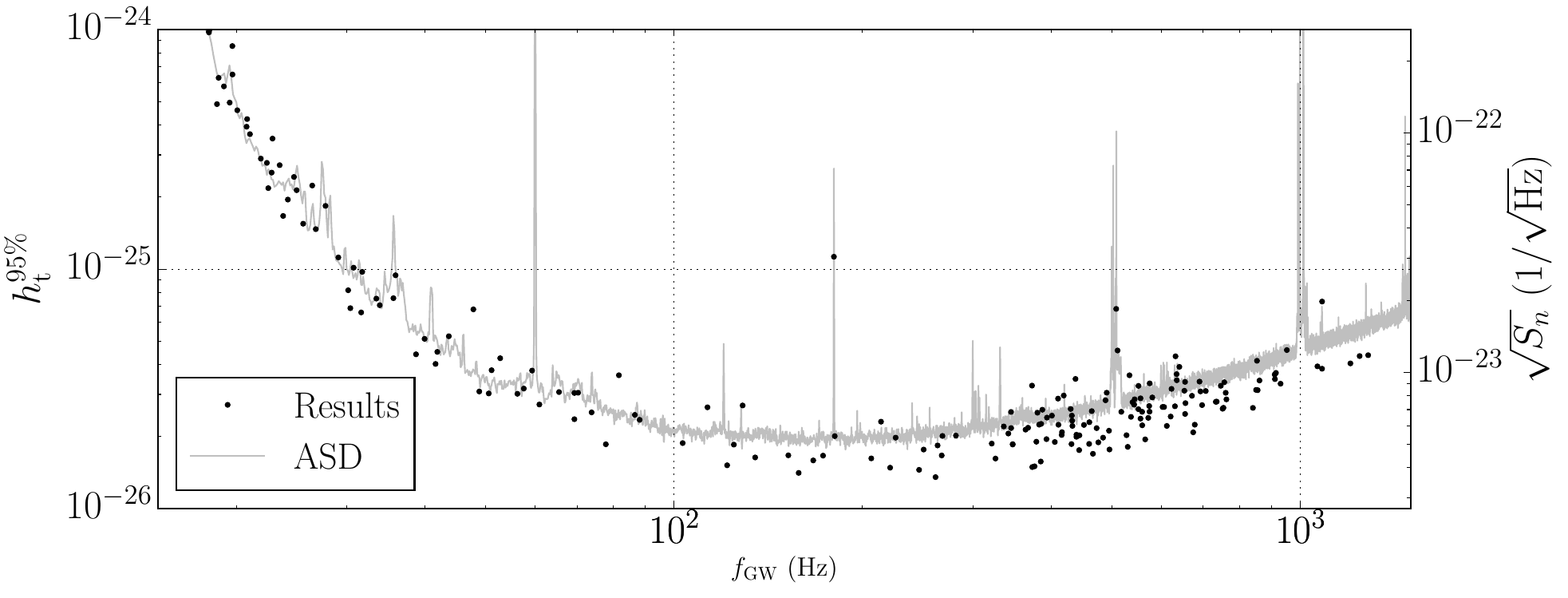}}%
\caption{{\em Non-GR upper limits vs emission frequency}. Circles mark the
95\%-credible upper limit on the scalar, $\hul{s}$ (top), and the effective
vector, $\hul{v}$ (middle), and tensor $\hul{t}$ (bottom) strain amplitudes as
a function of assumed gravitational-wave frequency for each of the \npsrs\ pulsars in our set.
The upper limits are obtained assuming a signal model including all five
independent polarizations ($\hyp{stv}$), and incorporating no information on
the orientation of the source (Table II in supplementary
material \cite{suppmat}). The effective amplitude spectral density (ASD) of the detector noise
is also displayed for reference; this is the harmonic mean of the H1 and L1
spectra; the scaling is obtained from linear regression to the upper limits.}
\label{fig:ul} \end{figure*}

\header{Conclusion.} We have presented the results of the first direct search
for nontensorial gravitational waves. This is also the first search
targeted at known pulsars that is sensitive to any of the
five measurable polarizations of the gravitational perturbation allowed by a generic metric
theory of gravity. From the analysis of O1 data from both aLIGO observatories,
we have found no evidence of signals from any of the \npsrs\ pulsars targeted.

In the absence of a clear signal, we have produced the first direct upper
limits for scalar and vector strains (\fig{ul}, and tables in the supplementary
material). The values of the 95\%-credible upper limits are comparable in
magnitude to previously-published GR constraints, reaching \red{$h\sim
1.5\times10^{-26}$} for pulsars whose frequency is in the most sensitive band
of our instruments.
This means that, to 95\% credibility, none of the pulsars in our set is emitting gravitational waves (tensorial or otherwise) at the frequencies analyzed with enough power for them to reach Earth with amplitudes larger than our upper limits.

Our results have been obtained in a theory-independent fashion. However, our
upper limits on nontensorial strain can be translated into model-dependent
constraints on beyond-GR theories by picking a specific alternative theory and emission mechanism. To
do so, one should use the upper limits produced under the assumption of a
signal model that incorporates the same polarizations allowed by the
theory one wishes to constrain; these may not necessarily be those in \fig{ul}
(e.g.\ for limits on a scalar-tensor theory, one needs upper limits from
$\hyp{st}$). However, this also requires nontrivial knowledge of the dynamics
of spinning neutron stars under the theory of interest.

While it is conventional to compare the sensitivity of continuous wave searches
to the canonical spin-down limit for each pulsar, it is not possible to do so
here without committing to a specific theory of gravity. This is because doing
so would require specific knowledge of how each polarization contributes to the
effective gravitational-wave stress-energy, how matter couples to the gravitational field, how
the waves propagate (dispersion and dissipation), and what the angular
dependence of the emission pattern is. However, analogues of the canonical
spin-down limit for specific theories may be obtained from the results
presented here by using the strain upper limits obtained assuming the
sub-hypotheses with polarizations corresponding to that theory, as mentioned
above.

We have demonstrated the robustness of searches for generalized polarization
states (tensor, vector, or scalar) in gravitational waves from spinning neutron
stars. Furthermore, even in the absence of a detection, we
were able to obtain novel constraints on the strain amplitude of nontensorial
polarizations. In the future, once a signal is detected, similar methods will
allow us to characterize the gravitational polarization content and, in so
doing, perform novel tests of general relativity. Although this search assumed
an emission frequency of twice the rotational frequency of the source,
this restriction will be relaxed in future analyses.

\begin{acknowledgments}

\header{Acknowledgments.} The authors gratefully acknowledge the support of
the United States National Science Foundation (NSF) for the construction and
operation of the LIGO Laboratory and Advanced LIGO as well as the Science and
Technology Facilities Council (STFC) of the United Kingdom, the
Max-Planck-Society (MPS), and the State of Niedersachsen/Germany for support of
the construction of Advanced LIGO and construction and operation of the GEO600
detector. Additional support for Advanced LIGO was provided by the Australian
Research Council. The authors gratefully acknowledge the Italian Istituto
Nazionale di Fisica Nucleare (INFN), the French Centre National de la Recherche
Scientifique (CNRS) and the Foundation for Fundamental Research on Matter
supported by the Netherlands Organisation for Scientific Research, for the
construction and operation of the Virgo detector and the creation and support
of the EGO consortium. The authors also gratefully acknowledge research support
from these agencies as well as by the Council of Scientific and Industrial
Research of India, Department of Science and Technology, India, Science \&
Engineering Research Board (SERB), India, Ministry of Human Resource
Development, India, the Spanish Ministerio de Econom\'ia y Competitividad, the
Conselleria d'Economia i Competitivitat and Conselleria d'Educaci\'o, Cultura i
Universitats of the Govern de les Illes Balears, the National Science Centre of
Poland, the European Commission, the Royal Society, the Scottish Funding
Council, the Scottish Universities Physics Alliance, the Hungarian Scientific
Research Fund (OTKA), the Lyon Institute of Origins (LIO), the National
Research Foundation of Korea, Industry Canada and the Province of Ontario
through the Ministry of Economic Development and Innovation, the Natural
Science and Engineering Research Council Canada, Canadian Institute for
Advanced Research, the Brazilian Ministry of Science, Technology, and
Innovation, Funda\c{c}ao de Amparo \`a Pesquisa do Estado de S\~{a}o Paulo
(FAPESP), Russian Foundation for Basic Research, the Leverhulme Trust, the
Research Corporation, Ministry of Science and Technology (MOST), Taiwan and the
Kavli Foundation. The authors gratefully acknowledge the support of the NSF,
STFC, MPS, INFN, CNRS and the State of Niedersachsen/Germany for provision of
computational resources.
This paper carries LIGO Document Number LIGO-P1700009.
\end{acknowledgments}

\appendix

\begin{ruledtabular}
\LTcapwidth=\textwidth
\begin{longtable*}{l|c|ccccccc|cc|ccc}
\caption{Results obtained using orientation information (see text for details).}
\label{tab:results_restricted}\\
\toprule
{} &   $f_{\rm GW}$ &  $\log {\cal B}^{\rm GR}_{\rm N}$ &  $~\log {\cal B}^{\rm GR+s}_{\rm N}$ &  $\log {\cal B}^{\rm GR+v}_{\rm N}$ &  $\log {\cal B}^{\rm GR+sv}_{\rm N}$ &  $\log {\cal B}^{\rm s}_{\rm N}$ &  $\log {\cal B}^{\rm v}_{\rm N}$ &  $~\log {\cal B}^{\rm sv}_{\rm N}$ &  $\log {\cal O}^{\rm S}_{\rm N}$ &  $\log {\cal O}^{\rm S}_{\rm I}$ & $h_{\rm s}^{95 \%}$ & $h_{\rm v}^{95 \%}$ & $h_{\rm t}^{95 \%}$\\
{} & {\tiny (Hz)} & {} & {} & {} & {} & {} & {} & {} & & {} & {} & {\tiny $\left(\times 10^{-26}\right)$} & {} \\ 
\midrule
\endfirsthead
\endhead
\midrule
\multicolumn{3}{r}{{Continued on next page}} \\
\midrule
\endfoot

\bottomrule
\endlastfoot
J0205+6449~~ & ~~30.41~~ &                                  -0.19 &                                    -0.07 &                                    -0.32 &                                     -0.26 &                                  0.06 &                                 -0.09 &                                  -0.10 &                                 -0.12 &  -0.53 &   14.79 &    7.99 &  3.70 \\
J0534+2200 & 59.33 &                                  -0.19 &                                    -0.40 &                                    -0.76 &                                     -0.88 &                                 -0.25 &                                 -0.55 &                                  -0.74 &                                 -0.47 &  -0.74 &    1.74 &    1.69 &  2.99 \\
J0835-4510 & 22.37 &                                  -0.06 &                                    -0.15 &                                    -0.29 &                                     -0.37 &                                 -0.07 &                                 -0.21 &                                  -0.26 &                                 -0.19 &  -0.67 &   16.66 &   13.78 & 15.27 \\
J1709-4429 & 19.51 &                                  -0.07 &                                    -0.02 &                                    -0.10 &                                     -0.07 &                                  0.01 &                                 -0.05 &                                  -0.04 &                                 -0.05 &  -0.60 &   61.33 &   36.31 & 34.34 \\
J1952+3252 & 50.59 &                                  -0.24 &                                    -0.43 &                                    -0.81 &                                     -0.98 &                                 -0.23 &                                 -0.58 &                                  -0.77 &                                 -0.50 &  -0.80 &    2.00 &    1.53 &  1.49 \\
\end{longtable*}

\end{ruledtabular}

\begin{ruledtabular}
\LTcapwidth=\textwidth
\begin{longtable*}{l|c|ccccccc|cc|ccc}
\caption{Results obtained without using orientation information (see text for
details).}
\label{tab:results}\\ 
\toprule
{} &     $f_{\rm GW}$ &  $~\log {\cal B}^{\rm s}_{\rm N}~$ &  $~\log {\cal B}^{\rm v}_{\rm N}~$ &  $~\log {\cal B}^{\rm t}_{\rm N}~$ &  $~\log {\cal B}^{\rm sv}_{\rm N}~$ &  $~\log {\cal B}^{\rm st}_{\rm N}~$ &  $~\log {\cal B}^{\rm vt}_{\rm N}~$ &  $~\log {\cal B}^{\rm svt}_{\rm N}~$ &  $\log {\cal O}^{\rm S}_{\rm N}$ &  $\log {\cal O}^{\rm S}_{\rm I}$ & $h_{\rm s}^{95 \%}$ & $h_{\rm v}^{95 \%}$ & $h_{\rm t}^{95 \%}$ \\
{} & {\tiny (Hz)} & {} & {} & {} & {} & {} & {} & {} & & {} & {} & {\tiny $\left(\times 10^{-26}\right)$} & {} \\ 

\midrule
\endhead
\midrule
\multicolumn{3}{r}{{Continued on next page}} \\
\midrule
\endfoot

\bottomrule
\endlastfoot
J0023+0923   &  655.69 &                                 -0.27 &                                 -0.57 &                                 -0.56 &                                  -0.76 &                                  -0.79 &                                  -1.05 &                                   -1.24 &                                 -0.65 &  -0.89 &    3.30 &    2.62 &    2.76 \\
J0024-7204AA & 1083.79 &                                  0.07 &                                 -0.27 &                                 -0.17 &                                  -0.21 &                                  -0.14 &                                  -0.47 &                                   -0.46 &                                 -0.20 &  -0.55 &   10.44 &    6.56 &    7.33 \\
J0024-7204AB &  539.86 &                                 -0.25 &                                 -0.55 &                                 -0.49 &                                  -0.72 &                                  -0.70 &                                  -0.93 &                                   -1.08 &                                 -0.60 &  -0.84 &    3.38 &    2.53 &    2.78 \\
J0024-7204C  &  347.42 &                                 -0.29 &                                 -0.68 &                                 -0.62 &                                  -0.86 &                                  -0.92 &                                  -1.24 &                                   -1.49 &                                 -0.73 &  -0.91 &    2.46 &    1.62 &    1.86 \\
J0024-7204D  &  373.30 &                                 -0.05 &                                 -0.32 &                                 -0.69 &                                  -0.28 &                                  -0.76 &                                  -0.89 &                                   -0.92 &                                 -0.45 &  -0.69 &    4.53 &    2.55 &    1.49 \\
\end{longtable*}

\end{ruledtabular}

\bibliography{gw,statistics}

\clearpage

\iftoggle{endauthorlist}{
 %
 %
 \let\author\myauthor
 \let\affiliation\myaffiliation
 \let\maketitle\mymaketitle
 \title{Authors}
 \pacs{}

 \newpage
 \maketitle
}

\end{document}